\documentclass[11pt]{article}

\usepackage{amsfonts,amsmath,amssymb,graphicx,bm}

\newcommand{\mycaption}[1]{\caption{\small #1}}

\newcommand{\bP}{\mathbf{P}}

\newcommand{\iopt}{\psi}
\newcommand{\mfac}{m}
\newcommand{\param}{\mathbf{p}}
\newcommand{\Ko}{K_o}
\newcommand{\To}{T_o}
\newcommand{\Ki}{\hat{A}}
\newcommand{\hd}{\hat{d}}
\newcommand{\Ash}{A^\sharp}
\newcommand{\Fstar}{F^*}
\newcommand{\Ups}{\Upsilon}
\newcommand{\Upsbar}{\overline{\Upsilon}}

\newcommand{\I}{\mathrm{i}}
\newcommand{\pr}{\mathbf{P}}
\newcommand{\ex}{\mathbf{E}}

\newcommand{\notthis}[1]{}

\newcommand{\indic}[1]{\mathbf{1}_{#1}}

\newcommand{\half}{\frac{1}{2}}

\newcommand{\cdl}{\,|\,}

\newcommand{\shalf}{{\textstyle\frac{1}{2}}}

\newcommand{\pderiv}[2]{\frac{\partial{#1}}{\partial{#2}}}

\begin{document}

\title{\bf Black to Negative\footnote{The title comes from the convention (at least in the UK) that on a battery, red is the positive terminal and black the negative. In accounting it is the other way round!} \\ \Large Embedded optionalities in commodities markets
}

\author{Richard J. Martin and Aldous Birchall}

\maketitle

\begin{abstract}
We address the modelling of commodities that are supposed to have positive price but, on account of a possible failure in the physical delivery mechanism, may turn out not to. This is done by explicitly incorporating a `delivery liability' option into the contract. As such it is a simple generalisation of the established Black model. 
\end{abstract}

\section*{Introduction}

To the consternation of many oil traders, WTI crude prices went below zero on 20th~April this year, with the May futures contract trading at $-37.63$ \$/bbl just before expiration. 
This prompted a rethink of the fundamentals of option pricing and risk management.

The standard pricing model is, of course, Black (also known as Black'76 or Black-Scholes), wherein the underlying asset price follows a geometric Brownian motion. As such a model cannot accommodate negative prices, there has predictably been a rush for the use of arithmetic (Bachelier) models instead, but this seems symptomatic of desperately clutching at the nearest simple probabilistic tool to hand, rather than of carefully appreciating the fundamental problem. Further, Black works well most of the time, albeit with the usual difficulties of pricing short-dated out-of-the-money options. It seems unwise simply to discard years of accumulated experience of Black in the following areas: hedging; pricing using volatility surfaces; the concept of convenience yield.

It is obvious from Table~\ref{tab:1} that the negative price was caused by a failure in the delivery mechanism. With storage tanks at Cushing full, and other means of storage such as transportation tankers in short supply, it was very difficult to store crude, and so holders of the May contract were in effect forced to pay people to take it off their hands. The deduction: physical commodities are not simply assets. What do we mean by this?

\begin{table}[!htbp]
\centering
\begin{tabular}{lrrrr}
\hline
& May & Jun & Jul & Aug \\
&	CLK0	& CLM0& CLN0 & CLQ0 \\
\hline 
17-Apr-20&	18.27 & 	25.03 & 	29.42  & 31.20 \\
20-Apr-20&	$-$37.63 & 	20.43 & 	26.28  & 28.51 \\
21-Apr-20&	10.01 & 11.57	& 18.69 &  21.61 \\
22-Apr-20&	- & 	13.78	& 20.69  & 23.76 \\
\hline
\end{tabular}
\mycaption{
Closing prices of recent WTI oil contracts just before and after the negative price date. Source: Bloomberg
}
\label{tab:1}
\end{table}

In \emph{Rich Dad, Poor Dad}, a series of `popular finance' books published over the last couple of decades, Robert Kiyosaki defines an asset as something that puts money in your pocket, whereas a liability takes it out. He goes on to point out that many things that people consider to be assets are in fact liabilities. A better summary would be: they are pure asset and a pure liability taken together. For example, a house has an intrinsic value, but on the liability side there is the cost of upkeep, and property taxes\footnote{We recall the apocryphal tale of the Russian hotel used as collateral against a bank loan. It was so dilapidated that the local authorities forced the new owner, i.e.\ the bank, to bring it to state of good repair---which cost more than the final value. Thereby the loan had negative recovery.}.

It follows that we should represent a commodity as a hypothetical pure asset $A$ say, whose value is always positive, coupled with a liability whose price is likely to be linked to $A$. Failure of the delivery mechanism can give rise to a large liability. It is attractive, in capital markets vernacular, to regard this as a form of intrinsic optionality that generally expires worthless---until one day it doesn't.
The option concept relates to previous research linking  storage cost and/or price volatility to inventory, which has traditionally been taken into account using the convenience yield \cite[\S2.2,2.3]{Geman05}. However, it goes further by identifying a problem that occurs when inventory becomes critical: the \emph{marginal} storage cost suddenly explodes and this can push the front-end futures negative. In extremis, these costs may not be limited to storage but could include unquantifiable legal liabilities arising from a breach of contract. In other words it is not simply storage cost, but cost arising from storage \emph{limitation}, that is being identified and captured.

Before making too strong a connection between commodities and securities markets, we should point out a fundamental  difference between them. In the latter, the spot price of the underlying asset has primacy, and the forward or futures price is linked to it by a simple parity argument based on either buying the asset and delivering it into the forward/futures, or selling it and buying it in the forward/futures market. In oil markets, however, there is no associated spot price to speak of: rather there are `benchmark prices' set by observations of the physical OTC spot market by agencies (known as Trade Journals) or by the futures market itself.
Consequently there is no notion of arbitrage\footnote{In its strict sense of a trade strategy providing a riskfree profit. The term is more loosely used by traders to describe what are more correctly called relative value trades.} between spot and futures or between one futures and another, because for such a trade to function one must be able to physically deliver into or receive physical delivery from the contract---the failure of which is the whole crux of the problem, given the complexity and expense of receiving barrels of crude oil rather than a conveniently dematerialised bond or share.

\section{Modifying the Black framework}
\label{sec:black}

\subsection{Pricing model}

See Bj\"ork \cite[Ch.20]{Bjork98} for information about martingale pricing of forward and futures contracts. Throughout $\ex$ means expectation under the risk-neutral measure $\bP$ (wherein with the rolled-up money market account as numeraire, discounted expectations are martingales), and $W_t$ is a Brownian motion under $\bP$. 
We are not using any subjective measure in this paper.

We write $A_t$ for the intrinsic asset price at time $t$. This is the value of the asset if we could store and deliver it at no cost, and is hypothetical.
Doing the obvious thing, we make the $A$-dynamics under $\bP$ a geometric Brownian motion, at least in the first instance:
\begin{equation}
dA_t/A_t = (r - y) \, dt + \sigma \, dW_t
\label{eq:At}
\end{equation}
where $r$ is the riskfree rate and $y$ is as usual the convenience yield \cite[Ch.2,3]{Geman05}.

Usually one has for the futures,
\[
F_t(T) = \ex_t[A_T],
\]
with a similar result holding for forwards\footnote{The expectation would be taken under the forward risk-neutral measure. As we are not worrying about stochastic interest rates in this paper, this difference will not detain us.},
but we are proposing
\begin{equation}
F_t(T) = \ex_t \left[ A_T - \iopt_-(A_T)  \right],
\label{eq:futs}
\end{equation}
where the intrinsic option $\iopt_-$ is a convex decreasing function. An obvious idea is a vanilla put option but as previously intimated we want the option price to rise without limit (explode) as $A_T\to0$. We also want to retain analytical tractability, as will be presently demonstrated. This points to a better idea, a call option on a negative power of $A_T$: 
\begin{equation}
\iopt_-(A_T) = \ell \Ki \big((\Ki/A_T)^\lambda - 1 \big)^+.
\label{eq:iopt-}
\end{equation}
There are three parameters, $\Ki>0$, $\ell\ge0$, $\lambda\ge0$: the first has the same units as $A_T$ and is interpreted as the price at which `things start to go wrong', and the other two are dimensionless and control the size of the resulting liability.

The quantities $\ell,\Ki,\lambda$ are allowed to depend on maturity, and therefore be different for different futures contracts. Delivery into the May contract is different from the June, etc, and so the optionalities relating to failure of the physical delivery mechanism may therefore vary substantially from one contract to another. Notice that there is no unique definition of the spot contract any longer, and this may seem odd. But as we have already noted, the spot oil contract in this context is a benchmark price based on observations of the spot OTC market which accommodates a multiplicity of contract specifications. As the futures market is far bigger than the spot market, we are modelling the futures directly rather than regarding it as the derivative of a spot contract and modelling the latter (as is common in capital markets).

\notthis{
We briefly return to a point we made earlier. It is superficially attractive to regard the following construct as a form of spot price:
\[
S_t = A_t - \iopt_-(A_t)
\]
so that $F_t(T)=\ex_t[S_T]$.
But there is a conceptual difficulty with that, as we may well need the parameters of $\iopt_-$ to dependent of the futures maturity $T$. This appears to generate a multiplicity of spot prices---but it \emph{is} legitimate, because in reality the spot contract does not trade.
}

The prescription (\ref{eq:iopt-}) has good analytical tractability in the sense that the expectation of $\iopt_-(A_T)$ can be evaluated using Black-type formulae:
\begin{eqnarray}
\ex_t[\iopt_-(A_T)] &=& \ell \Ki^{1+\lambda} \ex_t[A_T^{-\lambda}] \pr^{-\lambda}_t(A_T<\Ki) -\ell \Ki \pr_t(A_T<\Ki) \nonumber \\
&=& \ell \Ki \big( \mfac \Phi(-\hd_3) - \Phi(-\hd_2) \big)
\label{eq:exiopt} \nonumber
\end{eqnarray}
and so the futures price is
\begin{equation}
F = \Fstar - \ell \Ki \mfac \Phi(-\hd_3) + \ell \Ki \Phi(-\hd_2)
\label{eq:futpx}
\end{equation}
with
\begin{eqnarray}
\mfac &=& \left( \frac{\Ki e^{(\lambda+1)\sigma^2(T-t)/2}}{\Fstar} \right)^\lambda \nonumber \\
\hd_2 &=& \frac{\ln(\Fstar/\Ki) -\half\sigma^2(T-t) }{\sigma\sqrt{T-t}} \nonumber \\
\hd_3 &=& \hd_2 - \lambda \sigma\sqrt{T-t} \nonumber  \\
\Fstar &=&  \ex_t[A_T] = A_t e^{(r-y)(T-t)}  \nonumber
\end{eqnarray}
and $\Phi$ the standard Normal integral.
The quantity $\Fstar=\Fstar_t(T)$ is interpreted as the futures price on the intrinsic asset, were such a contract to trade; it is always positive even if $F$ is not.
The notation $\pr^\nu$ corresponds to the expectation $\ex^\nu[Z] \equiv \ex[A_T^\nu Z]/\ex[A_T^\nu]$, and we use standard arguments about change of num\'eraire.
This formula allows $\Fstar$ to be imputed from the futures price, but bear in mind that there is an implicit dependence on the specification of $\iopt_-$ and also on the volatility $\sigma$.

Valuation of European options on $F_t(T)$ can then be done using compound-option formulae. Let $\To\le T$ be the option maturity.
The main thing to note is that for any option strike $\Ko$ there is a unique $\Ash=\Ash(\Ko)$ such that
\[
\ex\big[ A_T - \iopt_-(A_T) \cdl A_{\To} = \Ash \big] = \Ko
\]
and so the exercise decision at time $\To$ reduces to  $A_{\To} \gtrless \Ash$.

As regards application we note that the exchange-traded options are vanilla options on each futures contract, with the same expiry\footnote{This is not quite correct: the options expire around five days before the futures. For this work we are setting $T$ equal to $\To$.} as the futures ($\To=T$). We deal only with this case here.
Also the exchange-traded options are American-style but we ignore the value of the early exercise option which, by standard arguments, is technically worthless.

The expected payouts of the call and put are
\begin{eqnarray}
C_{\Ko} &=& \ex_t[ (A_T-\iopt_-(A_T) - \Ko)^+ ] \nonumber \\
&=& \ex_t[ (A_T-\iopt_-(A_T) - \Ko) \indic{}(A_T>\Ash) ] \nonumber \\
&=& \ex_t[A_T] \pr^1_t(A_T>\Ash) - \Ko \pr_t(A_T>\Ash) \nonumber \\
&& \mbox{} - \ell \Ki \big[ \ex_t[(\Ki/A_T)^\lambda] \pr_t^{-\lambda}(\Ash < A_T < \Ki) - \pr_t(\Ash < A_T < \Ki) \big] \nonumber \\ 
&=& \Fstar \Phi(d_1) - \Ko \Phi(d_2) - \ell \Ki \cdot
\left\{ \begin{array}{ll} \mfac \big(\Phi(d_3)-\Phi(\hd_3)\big) -\big( \Phi(d_2) - \Phi(\hd_2) \big), & \Ash\le \Ki  \\
0, & \Ash\ge\Ki
\end{array} \right\} \label{eq:callpx}
\end{eqnarray}
and via the same route
\begin{eqnarray}
P_{\Ko} &=& \ex_t[ (\Ko - A_T + \iopt_-(A_T))^+ ] \nonumber \\
&=& \ex_t[ (\Ko - A_T + \iopt_-(A_T)) \indic{}(A_T<\Ash) ]   \nonumber \\
&=& \Ko \pr(A_T<\Ash) - \ex_t[A_T] \pr_t^1(A_T<\Ash) \nonumber \\
&& \mbox{} -\ell \Ki \, \ex_t\big[ \big( (\Ki/A_T)^\lambda-1\big) \indic{}(A_T<\Ki) \indic{}(A_T<\Ash) \big]  \nonumber \\
&=& \Ko \Phi(-d_2) - \Fstar \Phi(-d_1) + \ell \Ki \cdot
\left\{ \begin{array}{ll} 
\mfac \Phi(-d_3)-\Phi(-d_2)  , & \Ash\le\Ki \\
\mfac \Phi(-\hd_3)-\Phi(-\hd_2)  , & \Ash\ge\Ki
\end{array} \right\} \label{eq:putpx}
\end{eqnarray}
where 
\begin{eqnarray}
d_2 &=& \frac{\ln(\Fstar/\Ash) -\half\sigma^2(T-t) }{\sigma\sqrt{T-t}} \nonumber \\
d_1 &=& d_2 + \sigma\sqrt{T-t} \nonumber \\
d_3 &=& d_2 - \lambda \sigma\sqrt{T-t} \nonumber
\end{eqnarray}
and the others as before.
The prices are these discounted by $e^{-r(T-t)}$.
It is easy to see that the put-call parity formula
\[
C_{\Ko}-P_{\Ko} = F - \Ko
\]
is obeyed.

If $\To<T$ then the conditions $A_{\To}<\Ash$ and $A_T<\Ko$ refer to the intrinsic asset price at different times, and the expectations require the bivariate Normal integral. We will deal with this in forthcoming work.

We note for reference the elementary Black formula for the undiscounted call and put prices:
\begin{equation}
\begin{array}{rcl}
C_{\Ko} &=& F \Phi(d_+) - \Ko \Phi(d_-) \\ 
P_{\Ko} &=& \Ko \Phi(-d_-) - F \Phi(-d_+)
\end{array},
\qquad d_\pm = \frac{\ln(F/\Ko) \pm \half \sigma_B^2 (\To-t)}{\sigma_B\sqrt{\To-t}} .
\label{eq:black}
\end{equation}

\subsection{Numerical results}

An obvious approach to fitting is to minimise the root-mean-square relative pricing error with respect to the parameters $\sigma,\Ko,\lambda,\ell$:
\[
E = \left( \frac{1}{n} \sum_{j=1}^n \left( 
\frac{P^\star_{K_j}-P_{K_j}}{P^\star_{K_j}}
\right)^2 \right)^{1/2}
\]
($P^\star$ denotes the market price, $P$ the model).
Clearly at least four option prices will be needed, but this is not a difficulty as for oil they are quoted with strikes \$1 apart.
We have tried this for different contracts and on different dates and the results are consistently very good.

However, we should point out that different parameter sets can give similar fit, suggesting that four parameters is too many (we suspect that three would be ideal). To obviate this problem we can control the variation in intrinsic futures price $\Fstar$ between adjacent contracts. The reason for so doing is that we can make an inference about convenience yield, as we shall explain presently. The revised objective function is
\begin{equation}
E = \left( \frac{1}{n} \sum_{j=1}^n \left( 
\frac{P^\star_{K_j}-P_{K_j}}{P^\star_{K_j}}
\right)^2  + w^2 \sum_i  \left( \ln \frac{\Fstar(T_{i+1})}{\Fstar(T_i)}\right) ^2 \right)^{1/2}
\label{eq:Enew}
\end{equation}
with $w$ denoting a weighting coefficient (we use $w=5$). 

\begin{figure}[!htbp]
\begin{tabular}{ll}
(a) CLM0 \\
\scalebox{0.625}{\includegraphics*{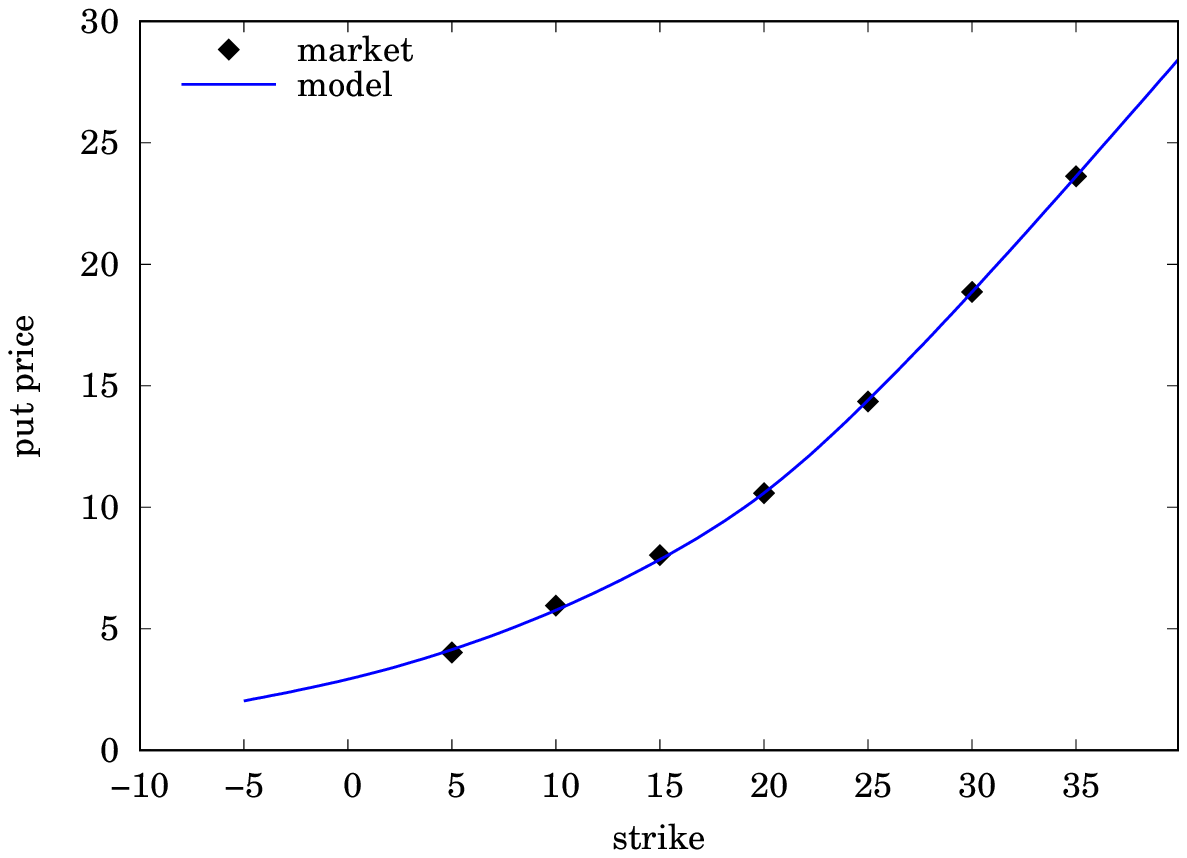}} &
\scalebox{0.625}{\includegraphics*{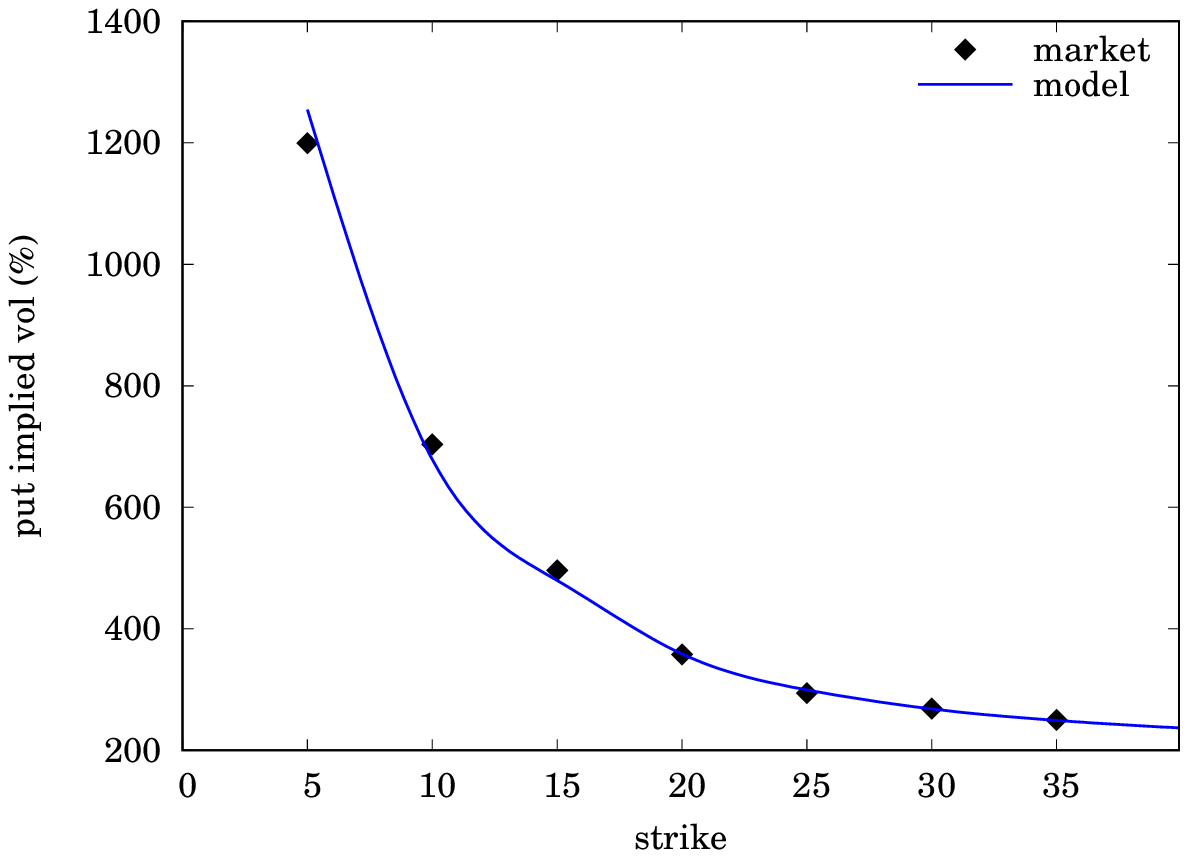}} \\
\\ \\
(b) CLN0 \\
\scalebox{0.625}{\includegraphics*{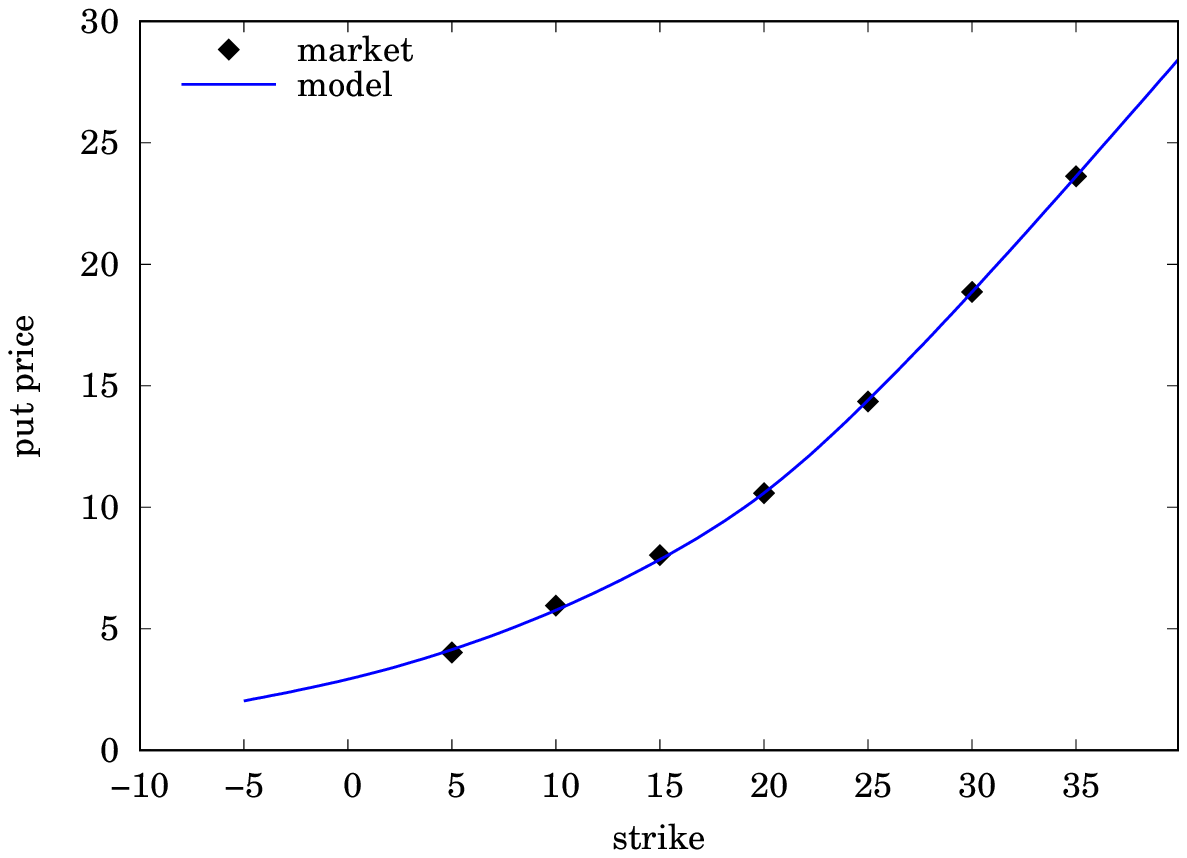}} &
\scalebox{0.625}{\includegraphics*{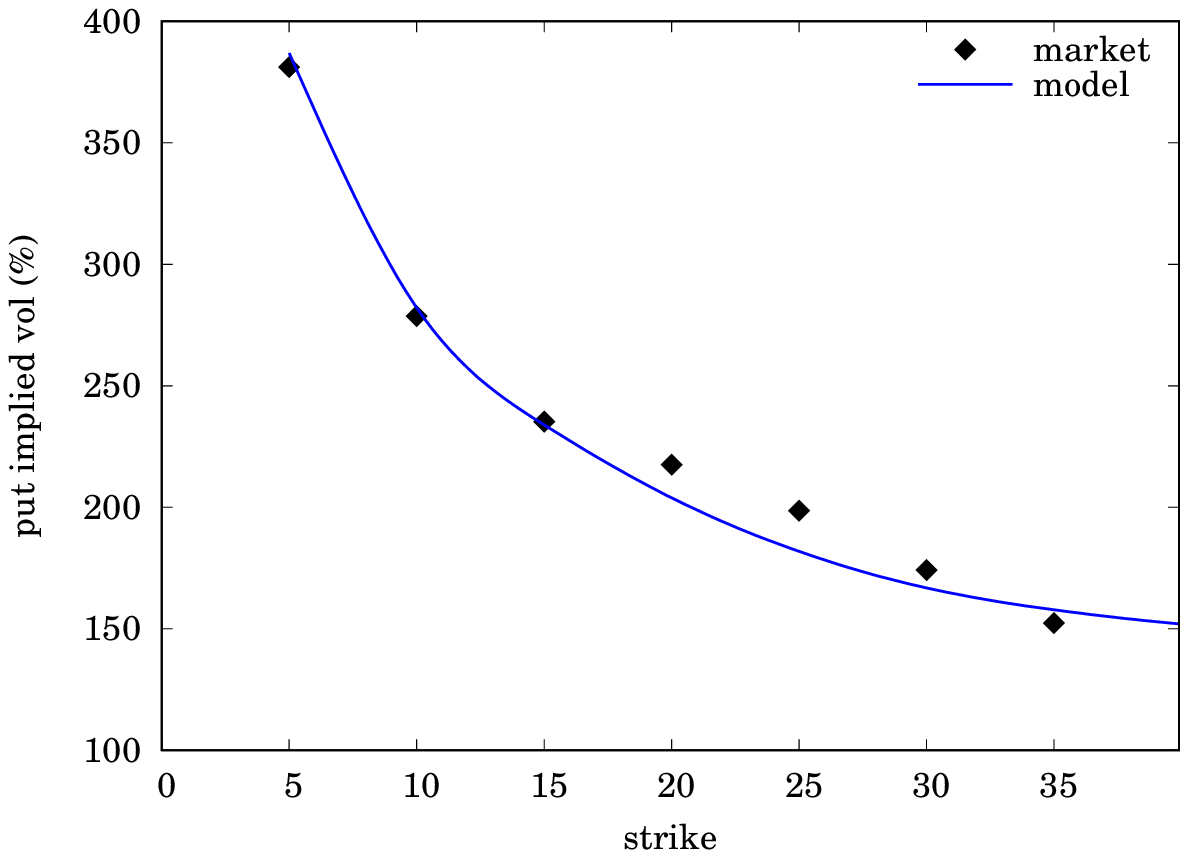}} \\
\\ \\
(c) CLQ0 \\
\scalebox{0.625}{\includegraphics*{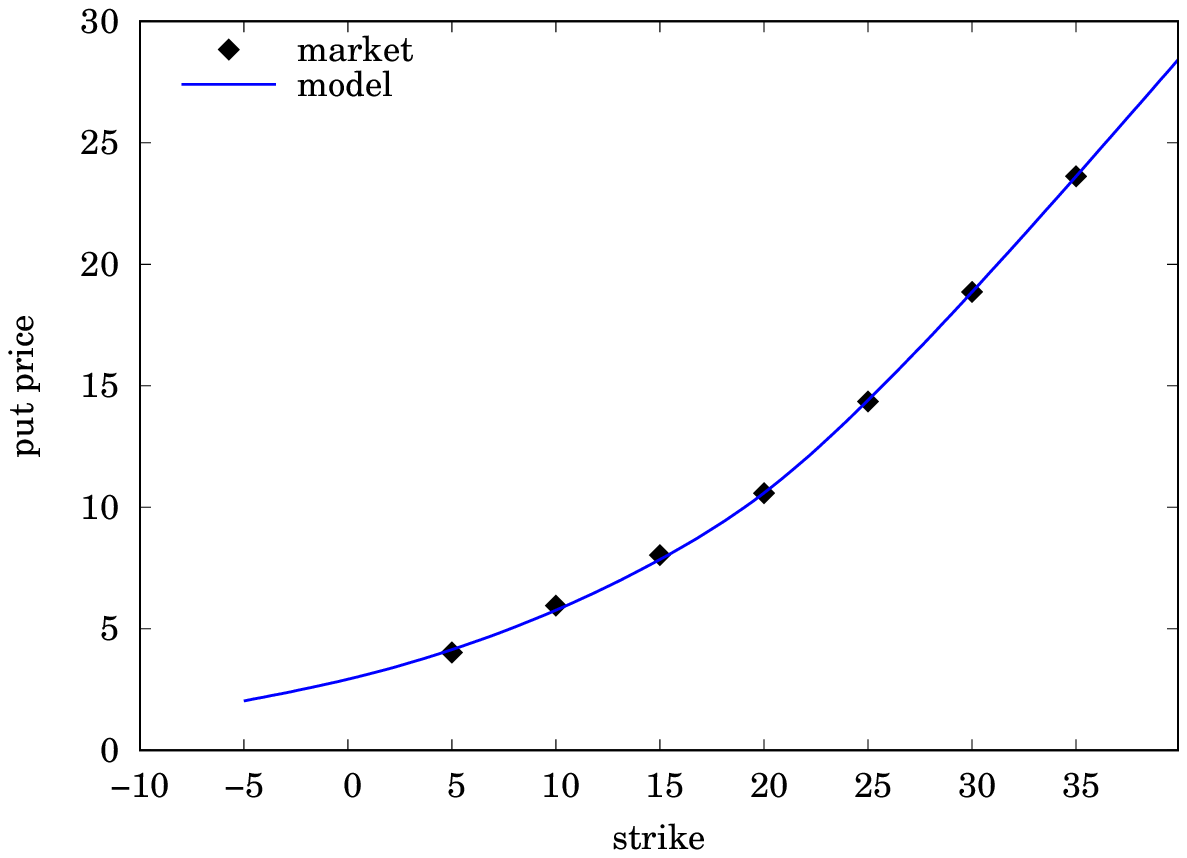}} &
\scalebox{0.625}{\includegraphics*{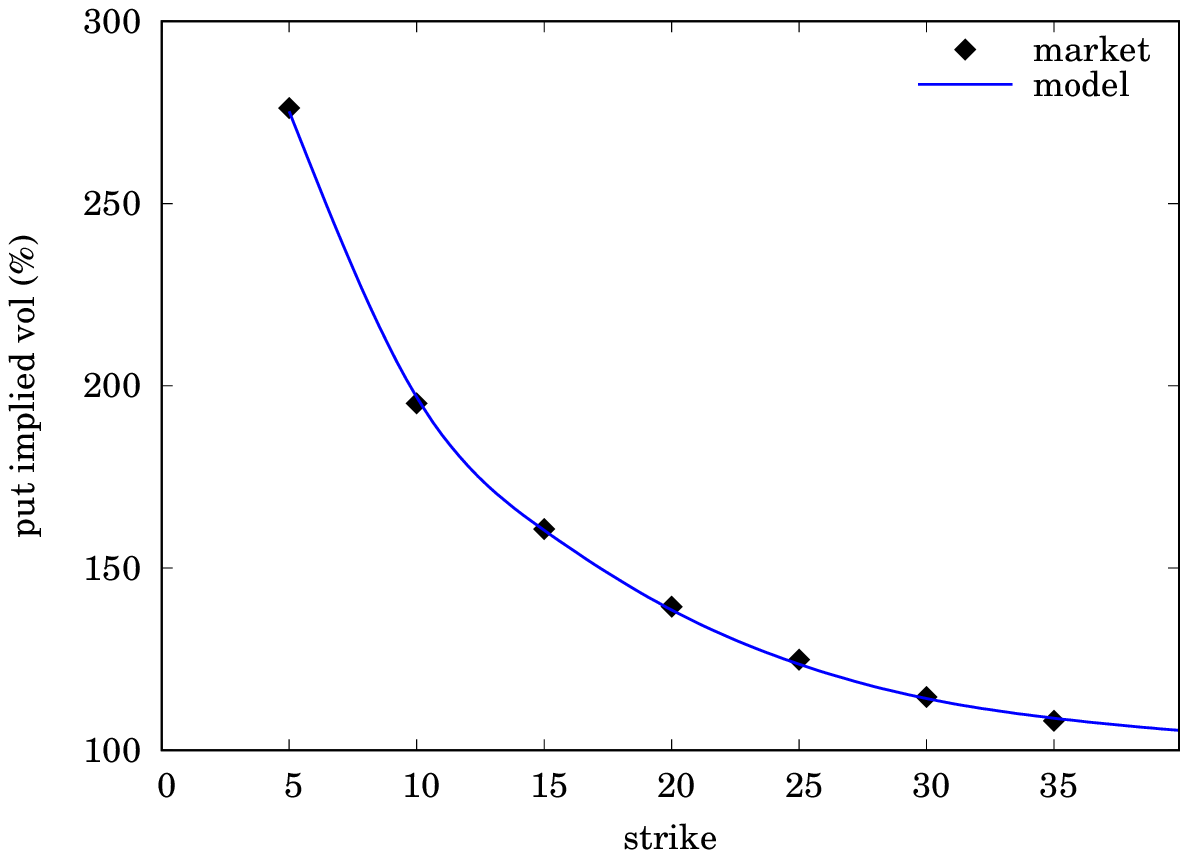}} \\
\end{tabular}
\mycaption{
Price and Black implied vol for front three CL contracts on 21-Apr-20: market and model compared. Market data source: Bloomberg.
}
\label{fig:1}
\end{figure}

As a specific example, Figure~\ref{fig:1} shows results for the June, July, August contracts CLM0, CLN0, CLQ0 as of 21-Apr-20.
The left-hand plot shows prices, and the right-hand plot the equivalent Black volatility (from eq.(\ref{eq:black}): obviously this can only be done for positive strikes).
It is apparent that the fit for low strikes is excellent.
And, for the first time ever, we can show a new Black model with negative strikes!
The r.m.s.\ relative pricing errors were $\approx 0.014,0.017,0.007$ for CLM0,CLN0,CLQ0 respectively, which are lower than the prevailing bid-offers. 

\begin{table}[!htbp]
\centering
\begin{tabular}{lr|rrrr|r}
\hline
& $F$ & $\sigma$ & $\Ki$ & $\lambda$ & $\ell$ & $\Fstar$ \\
\hline
CLM0 & 11.57 & 109\% & 21.7 & 0.921 & 2.20 & 20.42 \\
CLN0 & 18.69 & 125\% & 29.6 & 0.532 & 0.52 & 23.17 \\
CLQ0 & 21.61 & 88\% & 27.4 & 1.754 & 0.11 & 24.64 \\
\hline
\end{tabular}
\mycaption{
Fitted parameters for front three CL contracts on 21-Apr-20.
}
\label{tab:params1}
\end{table}

In all three cases the market quotes for the lowest-strike put options indicate positive probability of negative futures price at expiry, though obviously this is greatest for the front contract. Indeed, attempting to impose that the zero-strike put $P_0$ be worthless would lead to a convexity arbitrage in the puts (sell $2\times P_5$ and buy $1\times P_0$ and $1\times P_{10}$). The new model shows that $P_0$, and indeed some negative-strike puts, should have traded at a positive price, which is entirely reasonable.

Another important point is that the option prices are captured with parameters, shown in Table~\ref{tab:params1}, that have a reasonable physical intuition. While the volatility $\sigma$ is high (around 100\%), it retains some sort of plausibility: whereas an implied Black volatility $\sigma_B$ of over 1000\%, as is needed to mark the CLM0 5-strike put using (\ref{eq:black}), is essentially useless.

We return to the matter of convenience yield. Usually this is estimated from the ratio of adjacent futures prices $F$, but that method fails spectacularly when prices go negative. In view of the modelling herein, the correct approach is to use the `intrinsic futures' $\Fstar$ instead (this is necessarily positive), giving
\begin{equation}
\Fstar(T_i)/\Fstar(T_{i+1}) = e^{(r-y)(T_i-T_{i+1})}.
\label{eq:F-y}
\end{equation}
To obtain convenience yields with reasonable economic intuition we should not have extreme variation between adjacent $\Fstar$'s, and this is why (\ref{eq:Enew}) penalises that.
Applying this to the example shown\footnote{Note $T_2-T_1=\frac{34}{365}$ and $T_3-T_2=\frac{29}{365}$. We ignore $r$ as it is too small to be worth worrying about.} gives $y=-136\%$ as estimated from CLM0--CLN0 and $-88\%$ as estimated from CLN0--CLQ0. This indicates, as might be expected, extreme contango at the front end but less extreme further out.

A variation on eq.(\ref{eq:Enew}) is to require that $\sigma$ not vary too much between adjacent contracts---this idea naturally leads to term structure modelling, which we discuss in the next section. The model can be extended in a number of different directions, as the second half of the paper suggests. However, the main purpose of this paper is to show that the basic ideas hold good, even if the finer details of calibration are a matter for further research and development. Obviously given the closed-form nature of the option pricing, computation and therefore optimisation are very fast, which is an important advantage.

\subsection{Connection with the Merton model}

There are obvious parallels with the Merton structural credit model (i) in the sense of modelling assets and liabilities separately, with April~2020 being a sort of `credit event' in the oil market, and (ii) in that the introduction of an embedded option creates a volatility skew and may in other environments be an explanator of volatility skew/smile. 

In the Merton framework the equity $E_t$ of the firm is modelled as a call option on its assets $A_t$ and the debt as a riskfee bond minus a put option, struck at the face value of the firm's debt ($K$). 
By put-call parity, the equity is
\[
E_t = \ex_t[(A_T-K)^+] = A_t - \underbrace{\ex_t[\min(A_T,K)]}_{\mbox{\tiny embedded option}} .
\]
In the first instance let the firm value $A_t$ follow a geometric Brownian motion. The SDE followed by the equity $E_t$ is, by It\=o's lemma,
\begin{eqnarray*}
dE_t &=& r E_t \, dt + \Delta \sigma A_t \, dW_t \\
&=&  r E_t \, dt + \sigma_E(E_t) E_t \, dW_t
\end{eqnarray*}
where $\Delta$ is the call option's delta, and we find that $E_t$ has acquired local volatility given by
\[
\sigma_E = \sigma \Delta A_t / E_t.
\]
Now $E_t$ is a convex function of $A_t$, and so $\Delta > E_t/A_t$, and we conclude that $\sigma_E>\sigma$, the effect being less pronounced when the call option is in-the-money, corresponding to a firm with low leverage. Thus even if the firm value follows a geometric Brownian motion (which, from the perspective of calibrating to term structure of credit spread, is not a workable model) the equity price still acquires a volatility skew by virtue of the embedded option. A fuller discussion is in \cite{Martin07b}. More generally this connects with the idea of local volatility \cite{Dupire94}: rather than attempting to infer the local volatility surface from traded options, or imposing an arbitrary parametric form upon it, we may do better to instead postulate the existence of certain embedded optionalities, and identify those. As seen here, fixing up the Black model using local volatility would require an exorbitant level of volatility (even before worrying about the problems of negative price). On the other hand, incorporating an option into the traded asset seems to remove most of the difficulties, and the practicality is certainly enhanced by choosing a simple functional form, as here.


\section{Advanced skew/smile modelling}
\label{sec:more}

An embedded option generates a volatility skew/smile and the results of Figure~\ref{fig:1} show that in some cases it can explain most of the skew/smile. However, it is not necessarily true that all the volatility skew/smile is explainable that way. There are more obvious explanations, for example jumps. With that in mind, we extend the work in \S\ref{sec:black} to exponential L\'evy processes. But first we give a brief account of a simple and common construction and show what happens when we introduce an embedded option.

\subsection{Volatility surfaces}

As is well known, no single Black volatility prices all options consistently the market---in any asset class. This has led to a convenient visualisation tool: the (Black) volatility surface. 


We do not suggest that the incorporation of an `intrinsic put' of the type herein described will result in perfect calibration---but it does permit negative strikes and captures the `smile effect' for low strikes. For perfect matching we will need to make one parameter vary, and this had better be the volatility $\sigma$. With the Black model the implied volatility is unique (and exists provided the option price does not violate simple arbitrage constraints), because both call and put option prices are increasing functions of volatility. That the same property carries over to the model here is almost too obvious to be worth asking about, but there is a subtlety. When we alter $\sigma$, the calibrated futures price will change by (\ref{eq:futpx}) because the embedded option's value changes, and we will no longer match the market unless we alter the intrinsic asset price via a bump $\delta\Fstar$. Therefore in calibration when we talk about a move $\delta \sigma$ in volatility, we need also to apply a bump $\delta\Fstar$ so that the futures price is held fixed. The resulting sensitivity to $\sigma$ is not $\partial/\partial\sigma$ but rather
\begin{equation}
\left(  \pderiv{}{\sigma} \right)_F = 
 \pderiv{}{\sigma}  - \frac{\partial F/\partial \sigma}{\partial F /\partial \Fstar} \pderiv{}{\Fstar}.
\end{equation}
It is clear that for a call option, one has $(\partial C/\partial \sigma)_F > 0$ (as $\partial C/\partial\sigma>0$, $\partial C/\partial \Fstar>0$, $\partial F/\partial\sigma<0$, $\partial F/\partial \Fstar>0$). The same simple argument cannot be used for the put, but we can just use put-call parity instead. So finding implied volatility in the new model is straightforward.

\subsection{L\'evy models}

As mentioned earlier, these are the ideal tool for dealing with short-dated OTM options (for an introduction see e.g.~\cite{Schoutens03}), but exponential L\'evy models cannot deal with negative prices and/or strikes. Fortunately the work in \S\ref{sec:black} is easily generalised, as we now explain, first in a general setting and then with regard to a special case.

A L\'evy process is specified through its generator $L$, which is essentially a cumulant-generating function. We can deal either with the intrinsic asset price $A$ or with its corresponding futures $\Fstar$, and the latter is slightly neater\footnote{Because we do not need to invoke the convenience yield and risk-free rate at the outset, only to see them cancel out later. Put differently, $\Fstar$ must be a $\bP$-martingale and this imposes a condition on $L$.}.
Define $\ex_s[(\Fstar_t/\Fstar_s)^{\I u}] = e^{L(u) (t-s)}$, with $u$ a complex variable. The martingale condition is $L(-\I)=0$. 
For a Brownian motion, $L(u) = \mu \I u - \half \sigma^2 u^2$, and the martingale condition is $\mu=-\half \sigma^2$.

When jumps are included, $L$ acquires additional terms through jumps of size $y$ and intensity $\kappa(y)\ge0$, via
\[
L(u) = \mu \I u - \shalf \sigma^2 u^2 + \int_{-\infty}^\infty (e^{\I ux}-1) \kappa(x) \, dx;
\]
and in any L\'evy model of finite variation, $L$ may be cast in this form (L\'evy-Khinchin theorem).

Let us start with the basic Black model without the embedded optionality. Clearly the main challenge is calculating the expectation of $\indic{}(K_1<A_T<K_2)$ under $\pr^0$ and $\pr^1$.
The probability density of $x=\ln (\Fstar_T/\Fstar_t)$, under $\pr^0$, is obtained by Fourier transformation:
\[
f_T(x) = \frac{1}{2\pi} \int_{-\infty}^\infty e^{L(u)(T-t)- \I ux} \, du.
\]
This integral can be approximated by the discrete Fourier transform, followed by integration over $x\in[\ln(K_1/\Fstar_t),\ln(K_2/\Fstar_t)]$ to get the desired probability. We also need the probability under $\pr^1$, which corresponds to the exponentially-tilted density $e^{\nu x}f_T(x) \times \mbox{const}$, with $\nu=1$ in this instance. To this end define
\[
L^\nu (u) = L(u-\I \nu) - L(-\I \nu);
\]
then the density of $x$ under $\pr^\nu$ is obtained using the same integral as above, replacing $L$ with $L^\nu$.

When the embedded optionality is brought into play we need expectations of powers of $A_T$, which are given directly by $L$, and expressions of the form $\pr^\nu(K_1<A_T<K_2)$, which have already been dealt with above.
Therefore no extra machinery is required to calculate option prices in the new model.

Many choices of L\'evy model belong to so-called exponential families, by which is meant that if the generator $L(u,\param)$ is in the family ($\param$ denoting a list of parameters) then so is $L^\nu(u,\param)$. Equivalently, exponential tilting simply alters the parameters from $\param$ to $\param^\nu$, so that we may write $L^\nu(u,\param)=L(u,\param^\nu)$. Examples include the Variance Gamma and Normal Inverse Gaussian models, and the double-exponential and Merton jump-diffusions. Now for some families, the density $f_T$ and/or probability $\pr(K_1<A_T<K_2)$ are known in closed form. Write
\[
\pr_t(A_T>K)=\Ups\big(\ln(\Fstar/K),T-t;\param\big) \quad \mbox{and} \quad \Upsbar=1-\Ups
\]
where we recall that $\Fstar$ is short for $\Fstar_t(T)=\ex_t[A_T]$ and $\Fstar_T(T)=A_T$.
Then the (undiscounted) call and put prices are given by
\begin{eqnarray}
\begin{array}{rcl}
C_{K} &=& F_t \Ups\big(\ln(\Fstar_t/K),T-t;\param^1\big) - K \Ups\big(\ln(A_t/K),T-t;\param\big) \\[6pt]
P_{K} &=& K \Upsbar\big(\ln(\Fstar_t/K),T-t;\param\big) - F_t \Upsbar\big(\ln(A_t/K),T-t;\param^1\big),
\end{array}
\label{eq:levycp}
\end{eqnarray}
an obvious generalisation of (\ref{eq:black})\footnote{And in the Brownian case we note that if $\param=(\mu,\sigma)=(-\half\sigma^2,\sigma)$ then $\param^1=(\mu+\sigma^2,\sigma)=(\half\sigma^2,\sigma)$, and $\Upsilon(x,\tau,\param) = \Phi\big((x+\mu\tau)/\sigma\sqrt{\tau}\big)$, explaining the relationship between $d_+$ and $d_-$.}.

The introduction of embedded optionalities creates no problem, and (\ref{eq:futpx},\ref{eq:callpx},\ref{eq:putpx}) generalise via the following substitutions:
\begin{eqnarray*}
m &\rightsquigarrow& (\Ki/\Fstar)^\lambda e^{L(\I\lambda)(T-t)} \\
\Phi(d_1) &\rightsquigarrow& \Ups\big(\ln(\Fstar_t/\Ash),T-t;\param^{1}\big) \\
\Phi(d_2) &\rightsquigarrow& \Ups\big(\ln(\Fstar_t/\Ash),T-t;\param\big) \\
\Phi(d_3) &\rightsquigarrow& \Ups\big(\ln(\Fstar_t/\Ash),T-t;\param^{-\lambda}\big) \\
\Phi(\hd_2) &\rightsquigarrow& \Ups\big(\ln(\Fstar_t/\Ki),T-t;\param \big) \\
\Phi(\hd_3) &\rightsquigarrow& \Ups\big(\ln(\Fstar_t/\Ki),T-t;\param^{-\lambda}\big) .
\end{eqnarray*}

One concrete example is the double-exponential jump-diffusion \cite{Kou02}, in which exponentially-distributed positive and negative jumps of mean size $\xi_+$ and $-\xi_-$ arrive as independent Poisson processes of rate $\kappa_+$ and $\kappa_-$.
For this model
\[
L(u) = \mu \I u - \shalf \sigma^2 u^2 + \frac{\kappa_+\xi_+ \I u}{1 - \xi_+ \I u} - \frac{\kappa_-\xi_- \I u}{1 + \xi_- \I u} 
\]
with $\mu$ chosen so that $L(-\I)=0$, and if $\param=(\mu,\sigma,\kappa_+,\xi_+,\kappa_-,\xi_-)$ then
\[
\param^\nu = \left(\mu+\nu\sigma^2,\sigma,\frac{\kappa_+}{1-\nu\xi_+},\frac{\xi_+}{1-\nu\xi_+},\frac{\kappa_-}{1+\nu\xi_-},\frac{\xi_-}{1+\nu\xi_-}\right).
\]
Kou gives a closed-form expression for $\pr_t(A_T>K)$ as an infinite series of special functions (and in fact uses the symbol $\Ups$ for this purpose).

Another example is the Merton jump-diffusion (MJD) in which jumps are instead Normally distributed with mean $c$ and variance $\delta^2$, and arrive at rate $\kappa$.
The L\'evy generator is
\[
L(u) = (\mu-\shalf \sigma^2) \I u - \shalf \sigma^2 u^2 + \kappa\big(e^{c \I u -\delta^2 u^2/2} - 1\big)
\]
in which the martingale condition is $\mu=-\kappa(e^{c+\delta^2/2}-1)$, and if $\param=(\mu,\sigma,\kappa,c,\delta)$ then 
\[
\param^\nu = \big(\mu+\nu\sigma^2, \sigma, \kappa e^{\nu c+ \nu^2\delta^2/2}, c+\nu\delta^2, \delta\big).
\]
This can be handled using the general formulae above, but an alternative route to the same answer (the algebra is lengthy but not difficult) is to realise that it is simply a mixture of Black models, as is seen by conditioning on the number of jumps.
The MJD only provides a useful advance over a simple diffusion when the jumps are rare and large: very frequent small jumps are indistinguishable from a diffusion. Consequently this method of calculation is computationally effective as the sum can be truncated after only a few terms.

For the details, the probability of $j$ jumps in $[0,T-t]$ is $\pi_j = e^{-\theta}\theta^j/j!$, with $\theta=\kappa(T-t)$.
The futures price is now
\begin{equation}
F = \Fstar - \sum_{j=0}^\infty \pi_j \big( m_j \Phi(-\hd_{3,j}) - \Phi(-\hd_{2,j}) \big) 
\end{equation}
in which
\begin{eqnarray}
m_j &=&
\left( \frac{\Ki e^{(\lambda+1)\sigma^2(T-t)/2}}{\Fstar}  \right)^\lambda e^{ -\lambda \mu(T-t)  + \lambda j (-c+\lambda\delta^2/2) } \nonumber \\
\hd_{2,j} &=& \frac{\ln(\Fstar/\Ki) + (\mu-\half\sigma^2)(T-t)+jc}{\sqrt{\sigma^2(T-t)+j\delta^2}} \nonumber \\
\hd_{3,j} &=& \hd_{2,j} - \lambda \sqrt{\sigma^2(T-t)+j\delta^2}  \nonumber \; .
\end{eqnarray}
Moving on to the options, we have much as before
\begin{eqnarray}
C_{\Ko} &=& \sum_{j=0}^\infty \pi_j \biggr( \Fstar_j \Phi(d_{1,j}) - \Ko \Phi(d_{2,j})
\nonumber \\
&& \left. \mbox{} - \ell \Ki \cdot
\left\{ \begin{array}{ll} \mfac_j \big(\Phi(d_{3,j})-\Phi(\hd_{3,j})\big) -\big( \Phi(d_{2,j}) - \Phi(\hd_{2,j}) \big), & \Ash\le \Ki \\
0, & \Ash\ge\Ki 
\end{array} \right\} \right)
\\
P_{\Ko} &=& \sum_{j=0}^\infty \pi_j \biggr( \Ko \Phi(-d_{2,j}) - \Fstar_j \Phi(-d_{1,j}) 
\nonumber \\
&& \left. \mbox{} + \ell \Ki \cdot
\left\{ \begin{array}{ll} 
\mfac_j \Phi(-d_{3,j})-\Phi(-d_{2,j})  , & \Ash\le\Ki \\
\mfac_j \Phi(-\hd_{3,j})-\Phi(-\hd_{2,j})  , & \Ash\ge\Ki
\end{array} \right\} \right) 
\end{eqnarray}
where
\begin{eqnarray}
\Fstar_j &=& \Fstar e^{\mu(T-t)+jc+j\delta^2/2} \nonumber \\
d_{2,j} &=& \frac{\ln(\Fstar/\Ash) + (\mu-\half\sigma^2)(T-t)+jc}{\sqrt{\sigma^2(T-t)+j\delta^2}}
\nonumber \\
d_{1,j} &=& d_{2,j} + \sqrt{\sigma^2(T-t)+j\delta^2} \nonumber \\
d_{3,j} &=& d_{2,j} - \lambda \sqrt{\sigma^2(T-t)+j\delta^2}  \nonumber
\end{eqnarray}
(note that $\sum_{j=0}^\infty \pi_j\Fstar_j=\Fstar$ and $\sum_{j=0}^\infty \pi_j m_j=m$, recalling $\mu=-\kappa(e^{c+\delta^2/2}-1)$).

\subsection{Numerical results}

\begin{figure}[!htbp]
\begin{tabular}{ll}
(a) CLN0 \\
\scalebox{0.625}{\includegraphics*{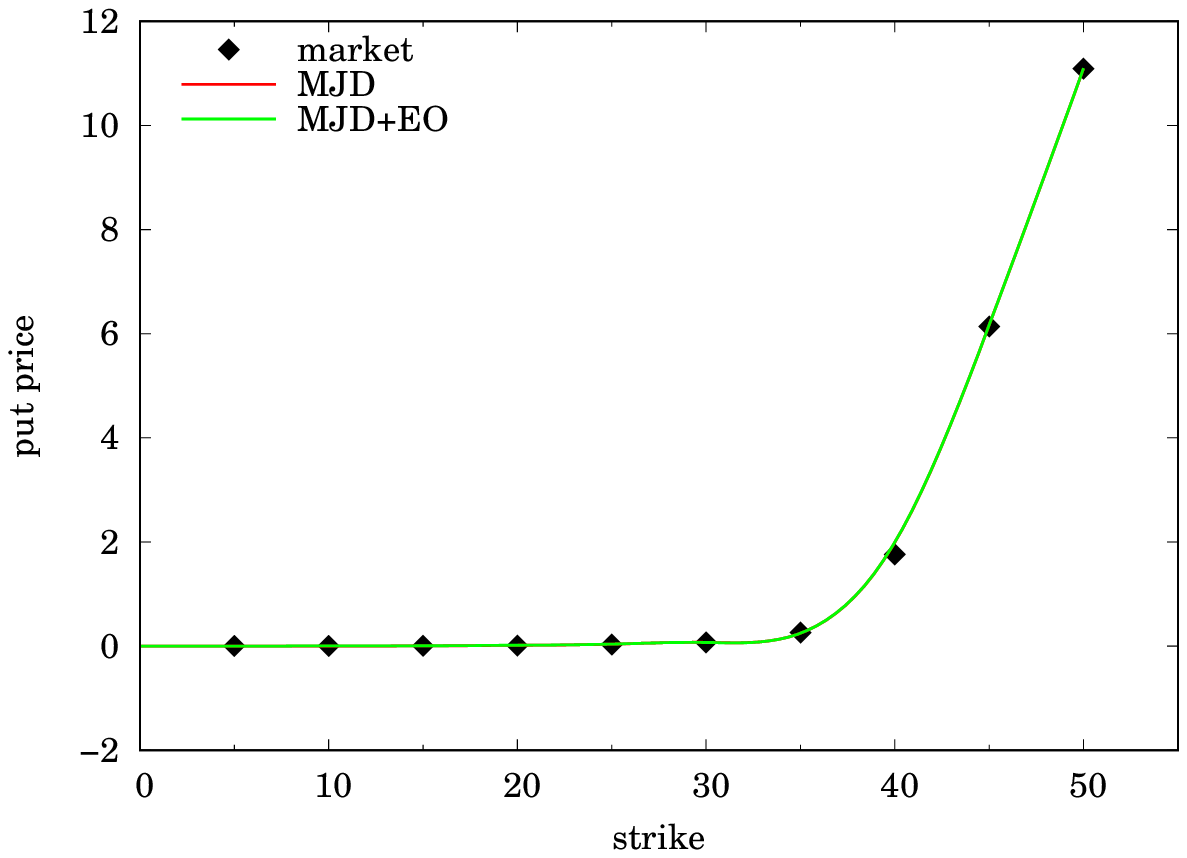}} &
\scalebox{0.625}{\includegraphics*{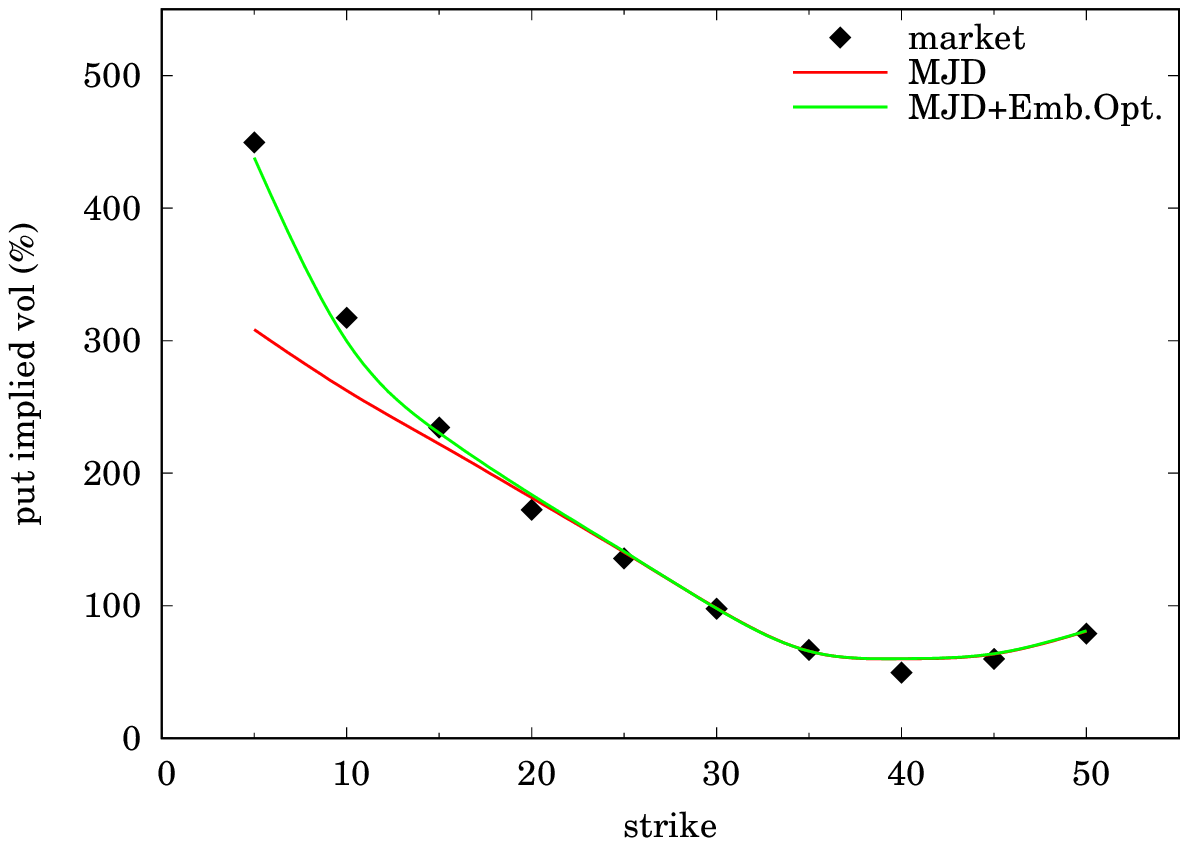}} \\
\\ \\
(b) CLQ0 \\
\scalebox{0.625}{\includegraphics*{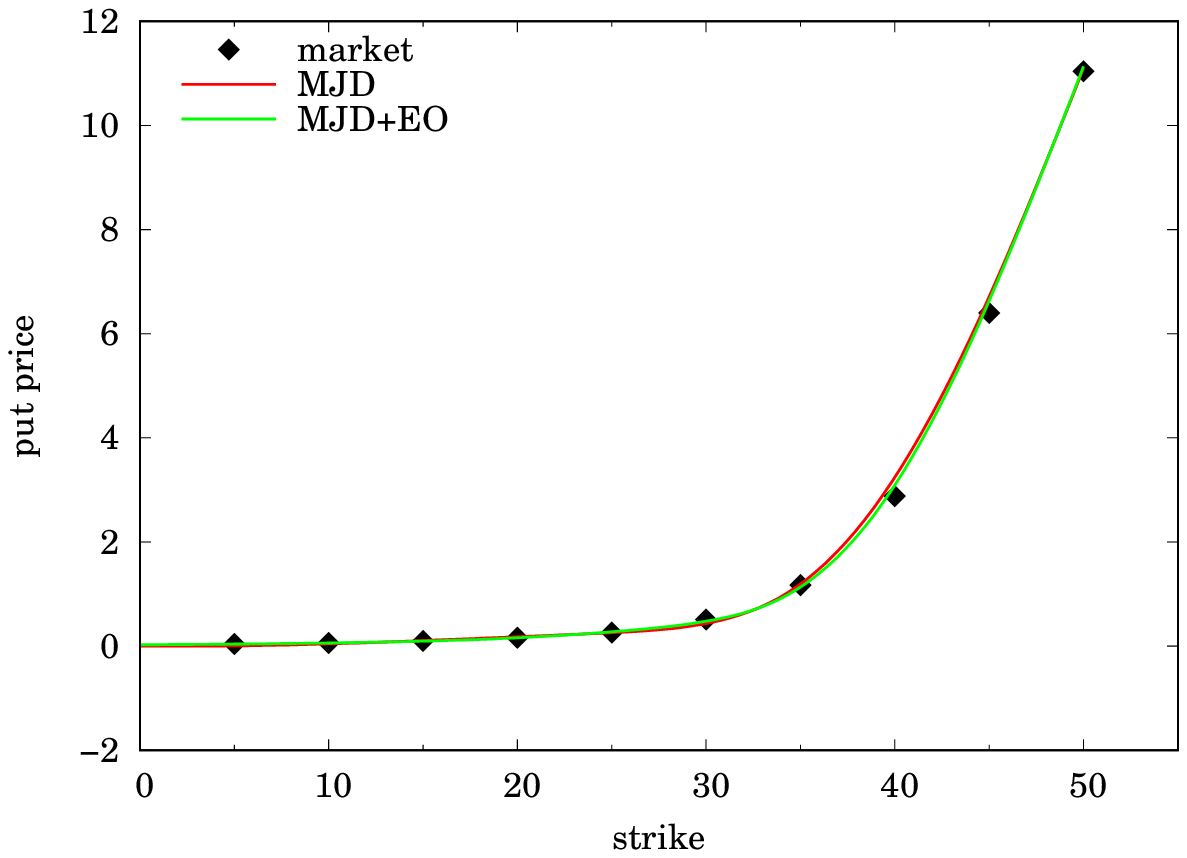}} &
\scalebox{0.625}{\includegraphics*{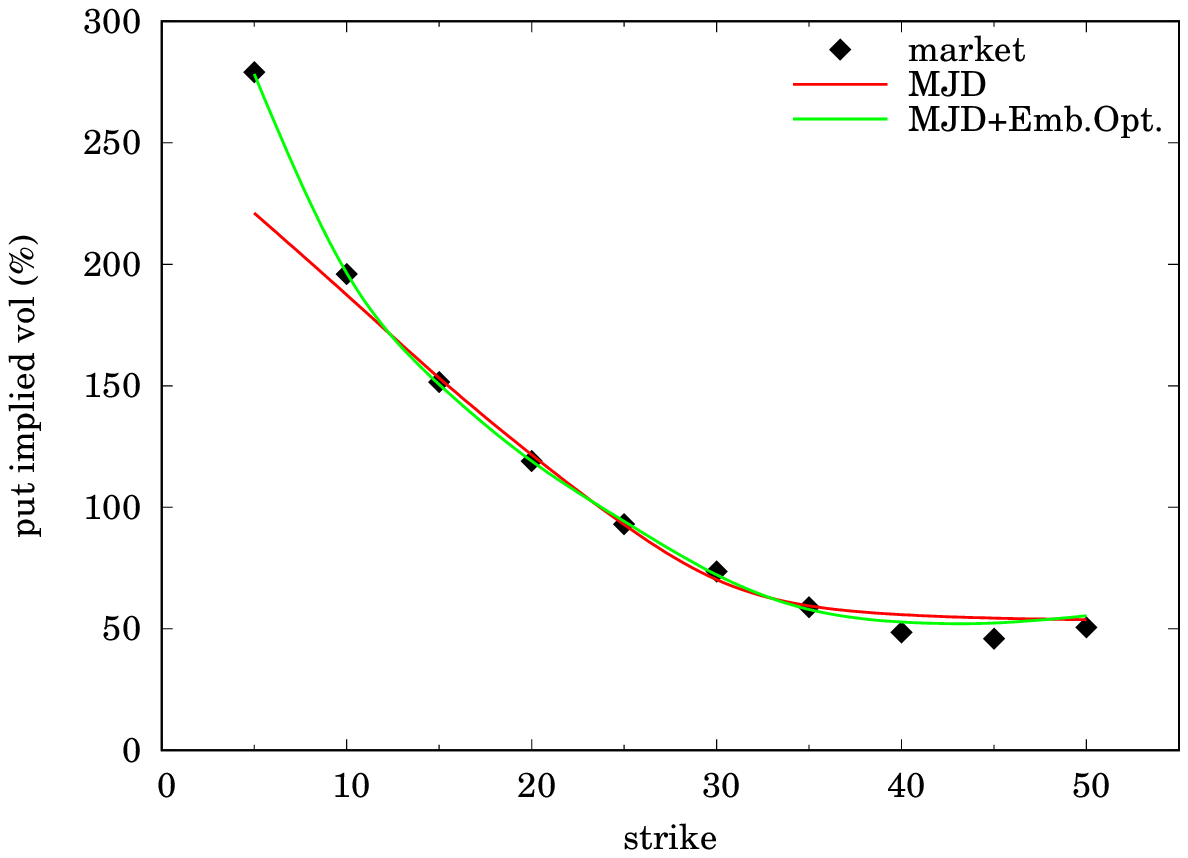}} \\
\\ \\
(c) CLU0 \\
\scalebox{0.625}{\includegraphics*{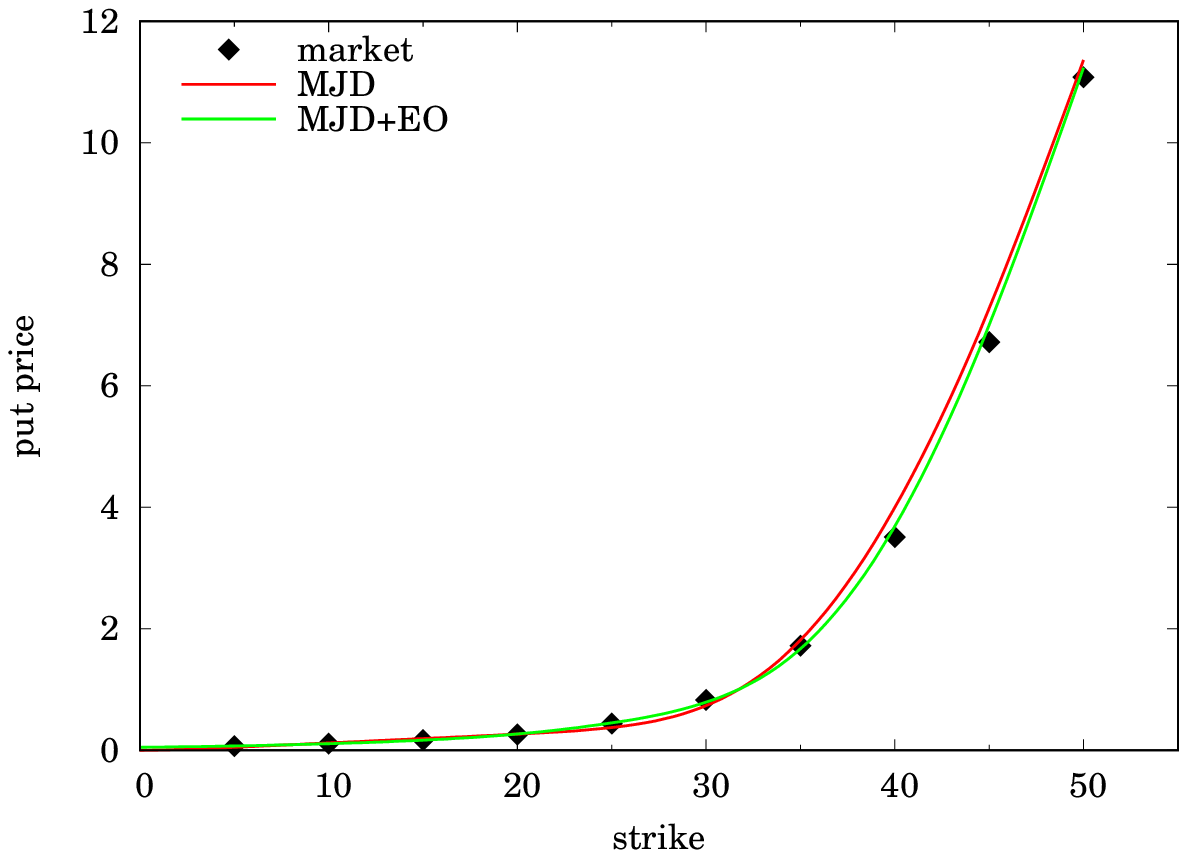}} &
\scalebox{0.625}{\includegraphics*{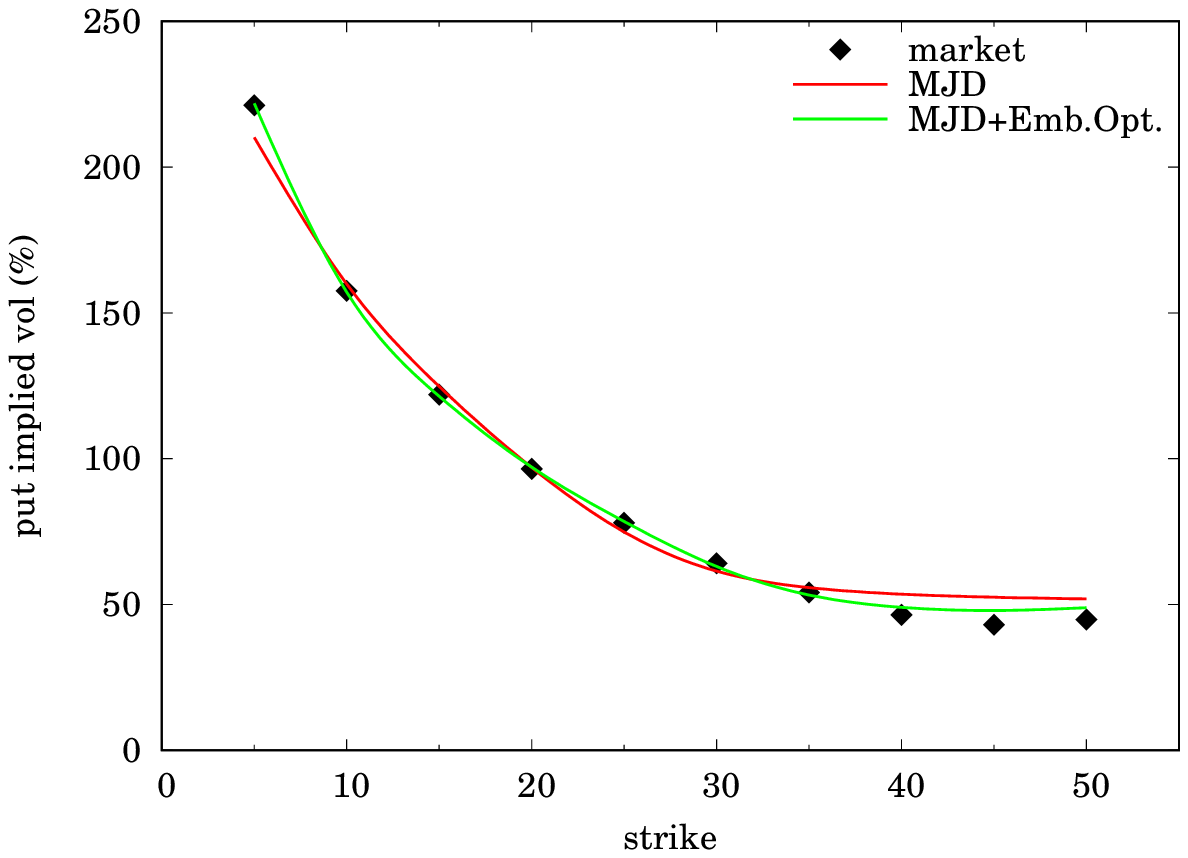}} \\
\end{tabular}
\mycaption{
Price and Black implied vol for front three CL contracts on 09-Jun-20: market and Merton jump-diffusion with and without embedded option compared. Market data source: Bloomberg.
}
\label{fig:2}
\end{figure}

As an example, we take the market a couple of months later, on 09-Jun-20. The oil price had now rallied to around 40. The results are shown in Fig.~\ref{fig:2}) with and without the embedded option, and the parameters are shown in Table~\ref{tab:params2}.
We have fixed $\delta=0.5$ throughout: this is to reduce the number of extra paremeters to two ($\kappa,c$) and to obviate the aforementioned degeneracy in the limit of very frequent tiny jumps.

The MJD alone fits well except for the lowest strikes, which the embedded option takes care of: the fitting error $E\approx 0.01$ in each case. The value of the embedded option, which is $\Fstar-F$, is much lower than in April. In `more normal' market conditions the skew is thereby explained, as one would expect: jumps capture the near-at-the-money volatility and the embedded option the deep-out-of-the-money volatility. The parameters do not vary wildly between contracts.

\begin{figure}[!htbp]
\begin{tabular}{ll}
(a) CLM0 \\
\scalebox{0.625}{\includegraphics*{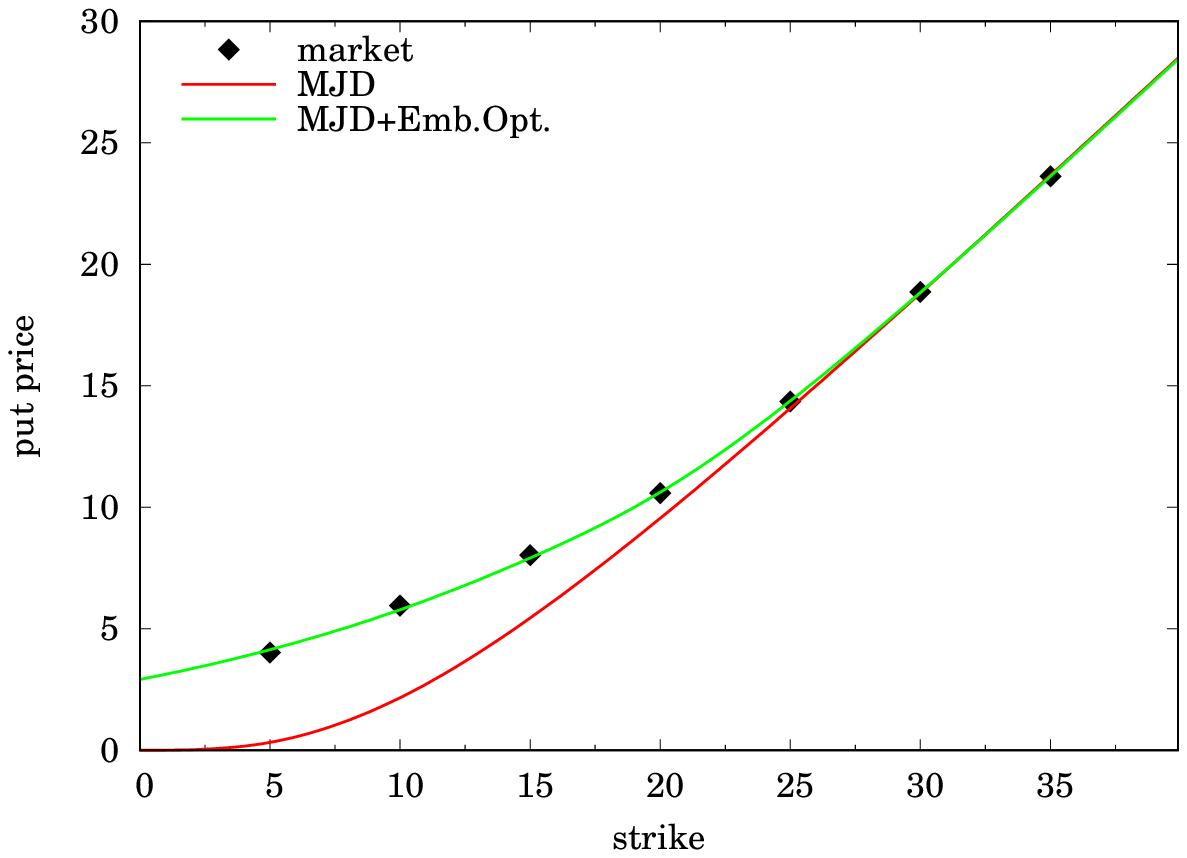}} &
\scalebox{0.625}{\includegraphics*{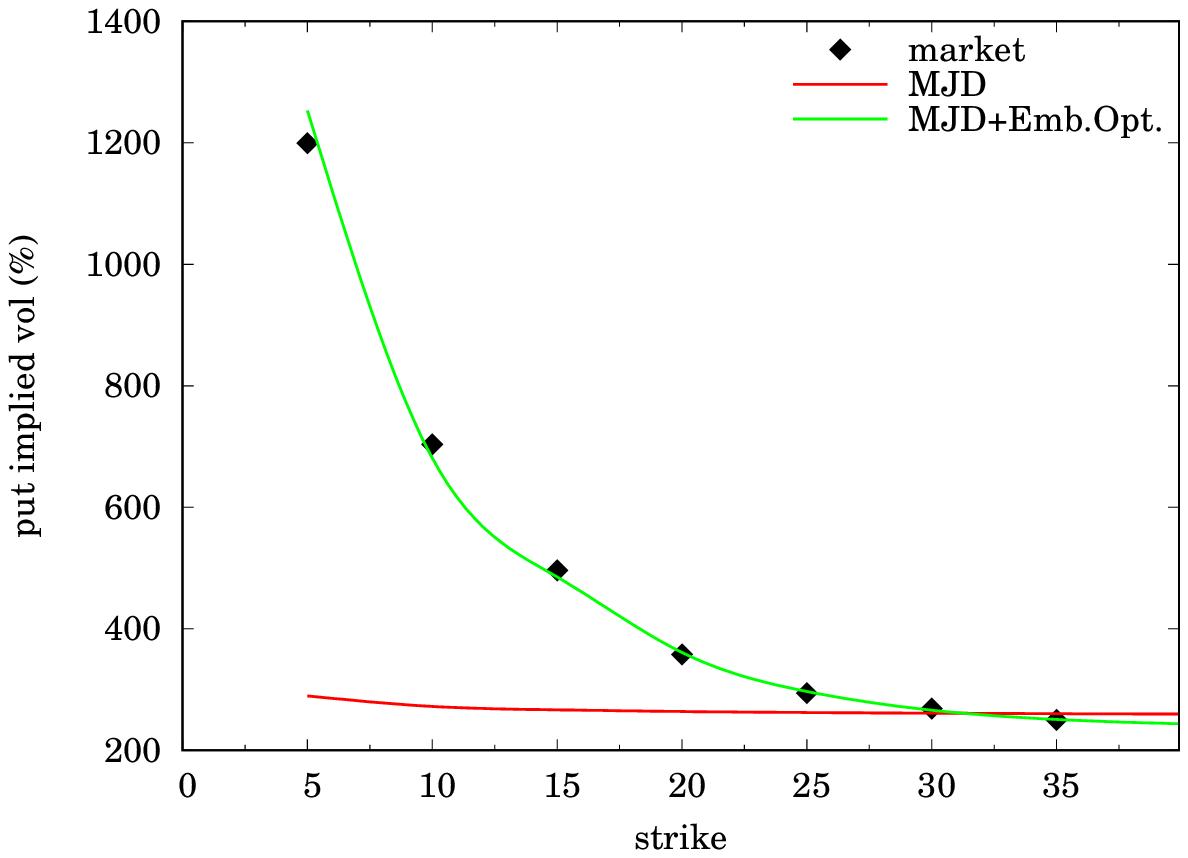}} \\
\\ \\
(b) CLN0 \\
\scalebox{0.625}{\includegraphics*{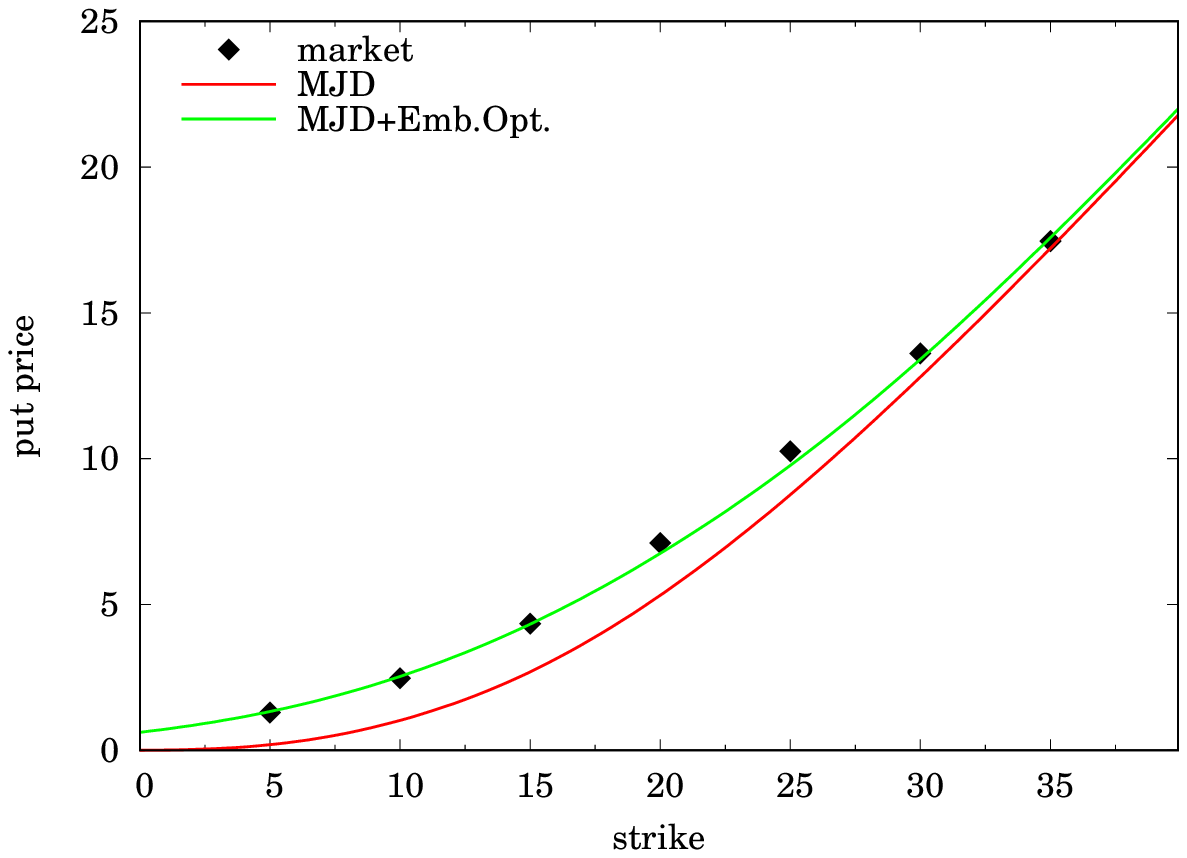}} &
\scalebox{0.625}{\includegraphics*{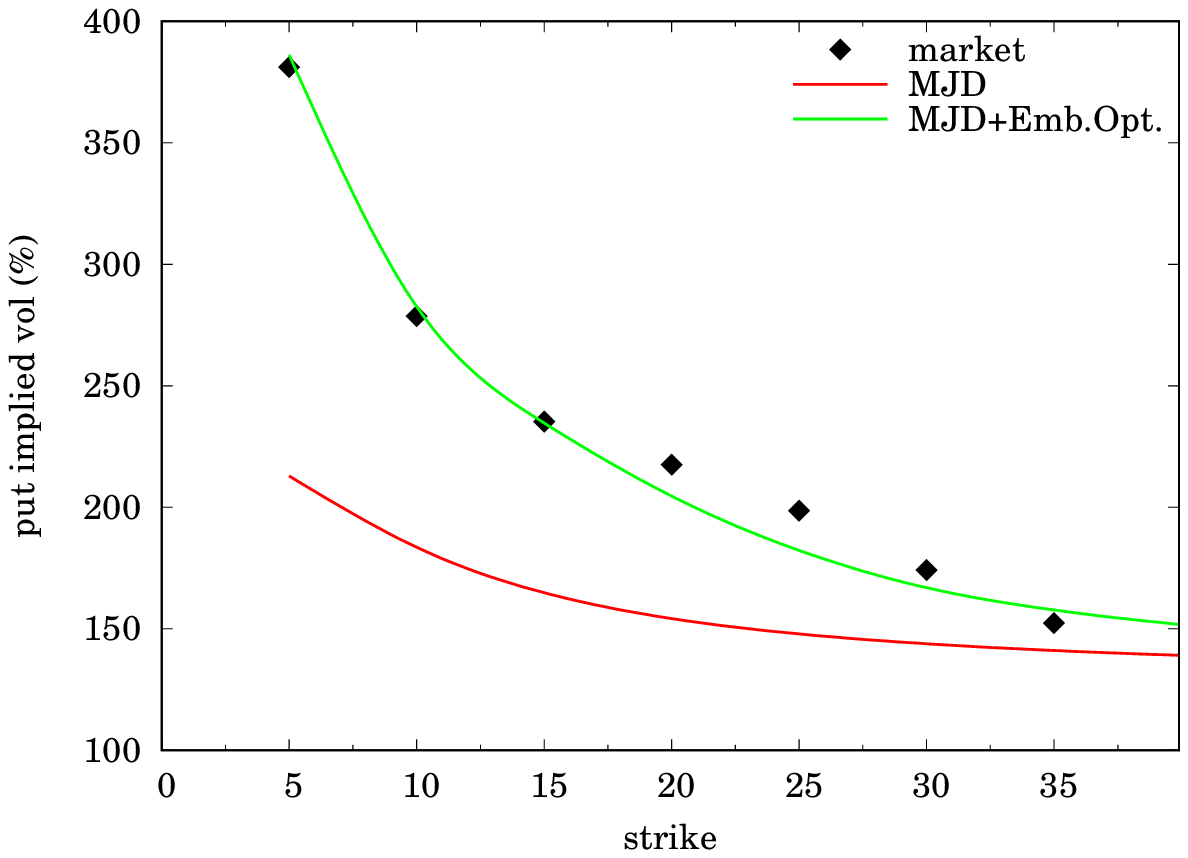}} \\
\\ \\
(c) CLQ0 \\
\scalebox{0.625}{\includegraphics*{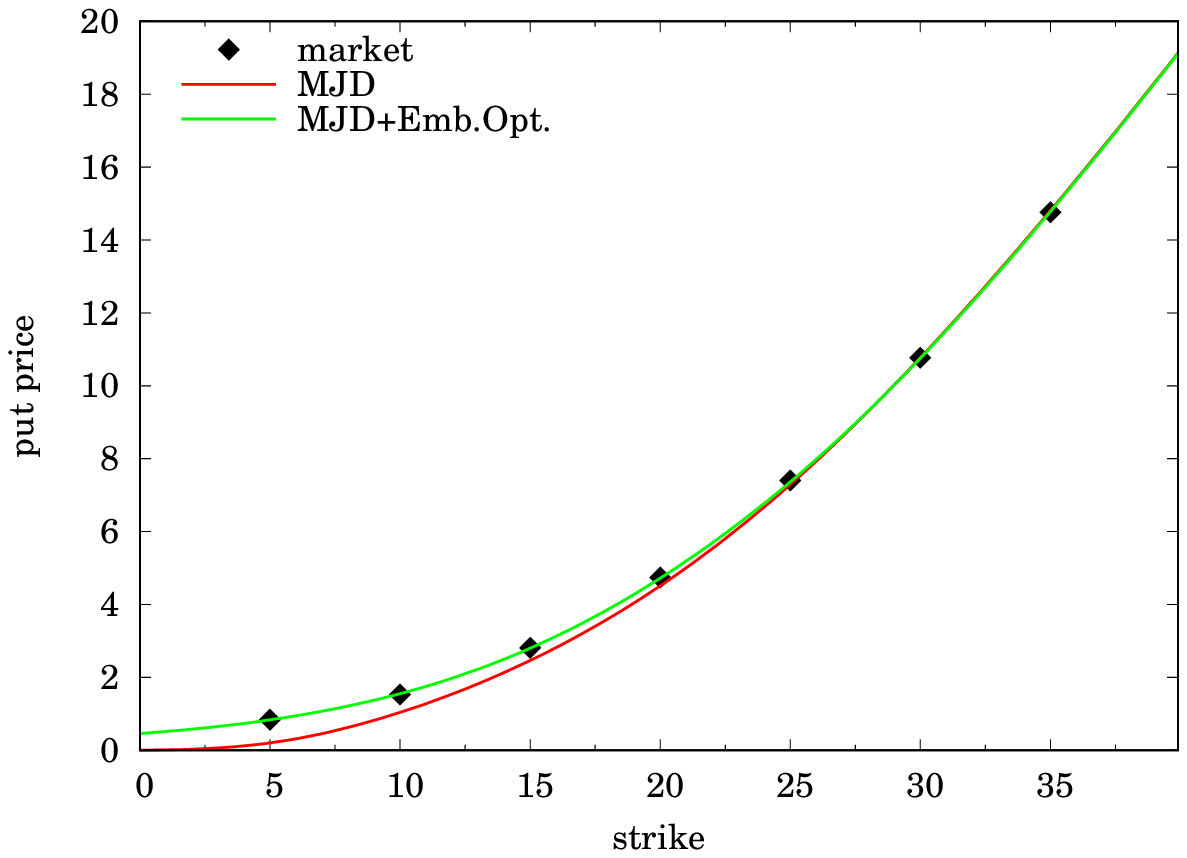}} &
\scalebox{0.625}{\includegraphics*{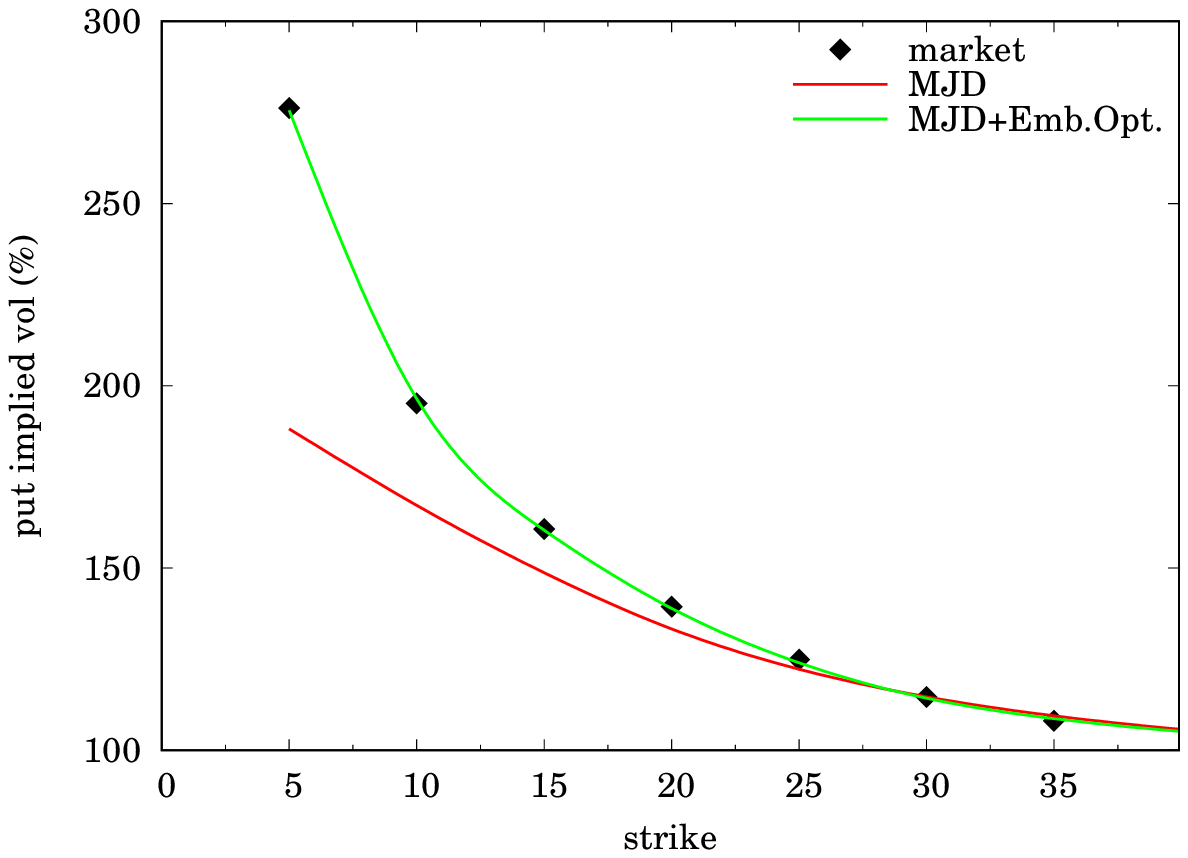}} \\
\end{tabular}
\mycaption{
Refit of 21-Apr-20 using MJD rather than Black (cf.~Fig.~\ref{fig:1}). A jump-diffusion on its own cannot generate enough skew (red trace), but an embedded option can. 
}
\label{fig:3}
\end{figure}

We can also redo the results from 21-Apr-20 using the MJD rather than a diffusion. In fact the incorporation of jumps makes no measurable difference to the fit---the green trace in Fig.~\ref{fig:3} is indistinguishable from the blue one in Fig.~\ref{fig:1}---showing that the entire smile/skew structure can be explained by the embedded option. The red trace in Figure~\ref{fig:3} shows that attempting to fit using only jumps, with no embedded option, fails completely. We repeat our previous comment that no exponential L\'evy model can deal with negative prices/strikes.

\begin{table}[!htbp]
\centering
\begin{tabular}{lr|rrrrrrr|r}
\hline
& $F$ & $\sigma$ & $\kappa$ & $c$ & $\delta$ & $\Ki$ & $\lambda$ & $\ell$ & $\Fstar$ \\
\hline
CLN0 & 38.94 & 56\% & 0.66 & $-0.30$ & 0.50 & 11.73 & 2.00 & 1.00 & 38.94 \\
CLQ0 & 39.16 & 44\% & 0.72 & $-0.32$ & 0.50 & 15.80 & 1.25 & 0.93 & 39.23 \\
CLU0 & 39.38 & 39\% & 0.61 & $-0.32$ & 0.50 & 15.82 & 1.17 & 0.99 & 39.50 \\
\hline
\end{tabular}
\mycaption{
Fitted parameters for front three CL contracts on 09-Jun-20.
}
\label{tab:params2}
\end{table}

\section{Convenience yield dynamics; Term structure}

Another direction in which the modelling can be generalised is in the modelling of convenience yield dynamics. 
This subject has received some attention over the years and an excellent account is given in \cite{Casassus05}, into which our work integrates completely.
A noticeable feature of commodity options is that the longer-dated futures are typically less volatile than the shorter-dated ones, and this suggests that the price is mean-reverting under $\bP$. With a nod to interest rate theory and in particular the Hull--White model \cite[\S17.11]{HullEd3}, it is obvious to consider the following general linear model, combining lognormal intrinsic asset prices with convenience yield dynamics and, optionally, interest rates (for the present purposes we neglect the last of these effects). Option prices can then be calculated in closed form using the methods of \S\ref{sec:black}.

By GL2, we mean a model in which the logarithm of the intrinsic asset price $x_t=\ln A_t$ and the convenience yield $y_t$ follow a bivariate Gaussian process of the form
\begin{equation}
\begin{bmatrix} dx_t \\ dy_t \end{bmatrix} = \Lambda 
\begin{bmatrix} x_t \\ y_t \end{bmatrix} \, dt +
 M(t)  \, dt +
\begin{bmatrix} \sigma_A \, dW^A_t \\  \sigma_y \, dW^y_t  \end{bmatrix}
\end{equation}
where the matrix $\Lambda$ is constant, and the drift term $M(t)$ and the volatilities $\sigma$ are not allowed to depend on $x$ or $y$ but may be time-dependent. By (\ref{eq:At}),
\begin{equation}
dx_t = (r-y_t -\sigma_A^2/2) \, dt + \sigma_A \, dW^A_t.
\label{eq:xt}
\end{equation}
The convenience yield follows essentially the Hull--White model, as also suggested by \cite{Gibson90}, see also \cite[Eq.(3.15)]{Geman05}, with an important difference in that it can be coupled to the asset price dynamics via a parameter $\beta\ge0$ thus:
\begin{equation}
dy_t = \kappa ( \alpha(t) + \beta x_t - y_t) \, dt + \sigma_y \,  dW^y_t.
\label{eq:yt}
\end{equation}
When $A_t$ is low, $y_t$ is more likely to be low/negative (contango) and when $A_t$ is high, $y_t$ is more likely to be positive (backwardation).
Thereby
\[
\Lambda = \begin{bmatrix} 0 & -1 \\ \kappa\beta & -\kappa \end{bmatrix}  , \qquad  M(t) = \begin{bmatrix} r-\sigma^2_A/2 \\ \kappa \alpha(t) \end{bmatrix} .
\]

The effect of the coupling parameter $\beta$ is to introduce an implicit mean reversion into the asset dynamics, because when the intrinsic asset price is low the futures curve is likely to be in contango and when high in backwardation.
The long-term variance is reduced, so that long-dated futures contracts are less volatile than short-dated ones. Models of this form therefore impose a term structure upon the futures price and also on the volatility, even if the parameters themselves are not maturity-dependent.

Backwardation and contango curves can be obtained easily enough in this model, the former by making $y_0$ and $\alpha(t)$ positive, the latter by making them negative. A humped curve is obtained by making $\alpha(t)>0$ but $y_0<0$. (We assume that the effect of $r$ is negligible.) At a simple level, we can argue that embedded optionalities just accentuate this. As an example, suppose that $y_0$ and $\alpha(t)$ are both negative, giving an upward-sloping curve, and suppose also that the parameters associated with $\iopt_-$ are not maturity-dependent. Then for the front contracts the embedded puts will be more in-the-money than the back ones, and so the modelled futures prices will be more strongly upward-sloping (as in April 2020). However, we repeat an important point we made earlier. The optionalities may well have different parameters for different maturity dates, and so in times of extreme stress the futures curve can in principle become contorted into weird shapes, as happens with the electricity market because storage in that market is next to impossible.

The obvious attraction of GL2 is that many quantities associated with it, principally the mean and variance of $A_T$, can be calculated in closed form. Indeed,
\[
\begin{bmatrix} x_t \\ y_t \end{bmatrix} =
\begin{bmatrix} x_0 \\ y_0 \end{bmatrix} +
\int_0^t e^{\Lambda (t-s)} \begin{bmatrix} r-\sigma_A^2/2 \\ \kappa \alpha(s) \end{bmatrix} \, ds +
\int_0^t e^{\Lambda(t-s)} \begin{bmatrix} \sigma_A \, dW^A_s \\ \sigma_y \, dW^y_s \end{bmatrix}
\]
and so the joint distribution of $x_t$ and $y_t$ is bivariate Normal with mean
\[
\begin{bmatrix} x_0 \\ y_0 \end{bmatrix}  + \int_0^t 
e^{\Lambda (t-s)} \begin{bmatrix} r-\sigma_A^2/2 \\ \kappa \alpha(s) \end{bmatrix} \, ds 
\]
and covariance matrix
\[
\int_0^t e^{\Lambda(t-s)} \begin{bmatrix} \sigma_A^2 & \rho \sigma_A \sigma_y  \\ \rho \sigma_A \sigma_y  & \sigma_y^2 \end{bmatrix} e^{\Lambda'(t-s)} \, ds ,
\] 
with $\rho$ the correlation between $dW^A_t$ and $dW^y_t$.
These expressions require matrix exponentiation, achieved by the following lemma\footnote{Proved by diagonalising the matrix.}: 
\begin{equation}
\exp\left( \begin{bmatrix} a & b \\ c & d \end{bmatrix} t \right) = e^{(a+d)t/2} \begin{bmatrix} 
\textstyle
\cosh \frac{\delta t}{2} + \frac{a-d}{\delta} \sinh \frac{\delta t}{2} & \frac{2b}{\delta} \sinh \frac{\delta t}{2} \\[6pt]
\frac{2c}{\delta} \sinh \frac{\delta t}{2} &
\cosh \frac{\delta t}{2} + \frac{d-a}{\delta} \sinh \frac{\delta t}{2}
\end{bmatrix}
\end{equation}
with  $\delta^2 = (a-d)^2+4bc$.

Contrary to what is implied in \cite{Casassus05} there is no requirement that the eigenvalues of $\Lambda$ be real: all that is necessary is that both have real part $\le0$, which is automatic provided that $\beta,\kappa\ge0$. In fact, when $\delta^2<0$, equivalent to $\beta>\kappa/4$, we will have oscillatory behaviour with period $2\pi\I/\delta$, as is observed in autoregressive processes with complex-conjugate poles.
It is worth noting that we could delete $\beta$ and make the mean reversion explicit, by adding a term $-\kappa_x x_t \, dt$ (where $\kappa_x \ge0$ is another parameter) into the drift term of (\ref{eq:xt}), replacing $0$ with $-\kappa_x$ in the top left-hand element of $\Lambda$. This has been suggested in e.g.~\cite[Eq.(3.8)]{Geman05}, but has a fundamentally different effect in that the eigenvalues of $\lambda$ must be real, so no cyclical behaviour can be generated.

The addition of stochastic interest rates via the Hull--White model gives us the GL3 model, and this presents no further difficulties in analysis provided the interest rate dynamics are not in any way driven by $A$ or $y$.
\notthis{
 which can be written in the same form as GL2:
\begin{equation}
\begin{bmatrix} dx_t \\ dy_t \\ dr_t \end{bmatrix} = \Lambda 
\begin{bmatrix} x_t \\ y_t \\ r_t \end{bmatrix} \, dt +
M(t) \, dt  +
\begin{bmatrix} \sigma_A \, dW^A_t \\  \sigma_y \, dW^y_t  \\ \sigma_r \, dW^r_t \end{bmatrix}
\end{equation}
Although the generator matrix $\Lambda$ is now $3\times3$, it can still be diagonalised easily enough. ***explain ***
} 

In this modelling framework we would fit all futures expiries (and hence their options) at once, as it is a term structure model. This makes the problem higher-dimensional but also imposes rigidity on the structure in the sense that the calibration parameters should not vary too strongly from one maturity to the next. Two restrictions are:
\begin{itemize}
\item[(i)]
One can no longer choose a different volatility for each maturity. The variation of volatility with $T$ is determined by $\sigma_A$ in (\ref{eq:xt}) and the parameters $\kappa$, $\beta$.
\item[(ii)]
As seen in (\ref{eq:F-y}), $\Fstar_t(T_i)/\Fstar_t(T_{i+1})$ gives an estimate of the convenience yield and so there cannot be too wild a variation in adjacent values of $\Fstar$.
\end{itemize}
Work on this area is continuing.
Regarding (ii), it is likely that this year's events will show a wide excursion of the convenience yield from its equilibrium level, adding to the catalogue of real-world examples in which the Ornstein--Uhlenbeck process fails to capture large deviations\footnote{A discussion of this and possible extensions to the OU model is given in \cite{Martin15b}.}.

\section{Conclusions and Final Remarks}

We have presented an extension of the Black model that incorporates an intrinsic optionality into the commodity price.  
It captures a liability caused by failure of the physical delivery process, and can cause negative futures prices, allowing the events of this April in the oil market to be captured with precision and insight. On the other hand in normal market conditions it constitutes only a small deformation from standard model frameworks.
It extends to exponential L\'evy models in a natural way and gives closed-form solutions for option pricing in the Merton jump-diffusion.


As an aside and opportunity for further work, it is not necessarily true that all optionalities have a negative impact on the price. Upward spikes in commodity prices can stem from the opposite kind of difficulty discussed here: low inventories, causing difficulty obtaining the physical asset. An obvious prescription in the light of this paper is an optionality of the form
\[
\iopt_+(A) = \ell A \big( (A/\Ki)^\lambda - 1 \big)^+
\]
where again $\ell,\lambda\ge 0$. This would in principle explain why in commodity markets implied volatilities often increase for high strikes, sometimes known as the `inverse leverage effect'.

\vspace{5mm}

\noindent
\emph{The opinions expressed in this paper are those of their authors rather than their institutions.
}

\noindent
\emph{Email {\tt richard.martin1@imperial.ac.uk}, {\tt aldous.birchall@trafigura.com}}

\notthis{
\section{Other bits}

If the option expiry is before the futures expiry then the position is more complicated: one has a compound option, the price of which can be evaluated in closed form using the bivariate Normal integral $\Phi_2$. It is worth noting that
\[
\ex \big[ \indic{X>a} \Phi(bX+c) \big] = \Phi_2 \left( \frac{\mu-a}{\sigma}, \frac{b\mu+c}{\sqrt{1+b^2\sigma^2}};\frac{b\sigma}{\sqrt{1+b^2\sigma^2}} \right)
\]
where $X\sim N(\mu,\sigma^2)$.
} 

\bibliographystyle{plain}
\bibliography{}

\begin{thebibliography}{10}

\bibitem{Bjork98}
T.~Bj\"ork.
\newblock {\em Arbitrage Theory in Continuous Time}.
\newblock Oxford University Press, 1998.

\bibitem{Casassus05}
J.~Casassus and P.~Collin-Dufresne.
\newblock Stochastic convenience yields implied from commodity futures and
  interest rates.
\newblock {\em J. Finance}, 60(5):2283--2331, 2005.

\bibitem{Dupire94}
B.~Dupire.
\newblock Pricing with a smile.
\newblock {\em RISK}, 7(1):18--20, 1994.

\bibitem{Geman05}
H.~Geman.
\newblock {\em Commodities and Commodity Derivatives}.
\newblock Wiley, 2005.

\bibitem{Gibson90}
R.~Gibson and E.~Schwartz.
\newblock Stochastic convenience yields and the pricing of oil contingent
  claims.
\newblock {\em J. Finance}, 45:959--976, 1990.

\bibitem{HullEd3}
J.~C. Hull.
\newblock {\em Options, Futures, and Other Derivatives}.
\newblock Prentice Hall, 1997.
\newblock 3rd ed.

\bibitem{Kou02}
S.~G. Kou.
\newblock A jump-diffusion model for option pricing.
\newblock {\em Mgt. Sci.}, 48(8):1086--1101, 2002.

\bibitem{Martin07b}
R.~J. Martin.
\newblock {CUSP} 2007: {A}n overview of our new structural model.
\newblock Technical report, Credit Suisse, 2007.
\newblock {\tt www.researchgate.net/profile/Richard\_Martin16}.

\bibitem{Martin15b}
R.~J. Martin, R.~V. Craster, and M.~J. Kearney.
\newblock Infinite product expansion of the {F}okker--{P}lanck equation with
  steady-state solution.
\newblock {\em Proc. R. Soc. A}, 471(2179):20150084, 2015.

\bibitem{Schoutens03}
W.~Schoutens.
\newblock {\em L\'evy Processes in Finance}.
\newblock Wiley, 2003.

\end{thebibliography}

\clearpage

\end{document}